\documentclass[apj]{emulateapj}
\usepackage{natbib, aas_macros}
\newcommand\feii{Fe\,{\footnotesize II}}
\newcommand\mgii{Mg\,{\footnotesize II}}

\shorttitle{Chemical evolution of the Universe derived from the BLR of
quasars} \shortauthors{Sameshima, Yoshii \& Kawara}

\begin{document}

\title{Chemical evolution of the Universe at $0.7 < z < 1.6$ derived
 from abundance diagnostics of the broad-line region of quasars}

\author{H. Sameshima\altaffilmark{1}, Y. Yoshii\altaffilmark{2} and K. Kawara\altaffilmark{2,3}} 

\altaffiltext{1}{Laboratory of Infrared High-resolution spectroscopy
(LIH), Koyama Astronomical Observatory, Kyoto Sangyo University,
Motoyama, Kamigamo, Kita-ku, Kyoto 603-8555, Japan. E-mail:
sameshima@cc.kyoto-su.ac.jp} \altaffiltext{2}{Institute of Astronomy,
School of Science, University of Tokyo, 2-21-1 Osawa, Mitaka, Tokyo
181-0015, Japan} \altaffiltext{3}{Deceased January 2015.}

\begin{abstract}
 We present an analysis of \ion{Mg}{2} $\lambda2798$ and \ion{Fe}{2} UV
 emission lines for archival Sloan Digital Sky Survey (SDSS) quasars to
 explore diagnostics of the magnesium-to-iron abundance ratio in a
 broad-line region cloud. Our sample consists of 17,432 quasars selected
 from the SDSS Data Release 7 with a redshift range of $0.72 < z <
 1.63$. A strong anticorrelation between \ion{Mg}{2} equivalent width
 (EW) and the Eddington ratio is found, while only a weak positive
 correlation is found between \ion{Fe}{2} EW and the Eddington ratio. To
 investigate the origin of these differing behaviors of \ion{Mg}{2} and
 \ion{Fe}{2} emission lines, we have performed photoionization
 calculations using the {\sc Cloudy} code, where constraints from recent
 reverberation mapping studies are considered.  We find from
 calculations that (i) \ion{Mg}{2} and \ion{Fe}{2} emission lines are
 created at different regions in a photoionized cloud, and (ii) their EW
 correlations with the Eddington ratio can be explained by just changing
 the cloud gas density. These results indicate that the
 \ion{Mg}{2}/\ion{Fe}{2} flux ratio, which has been used as a
 first-order proxy for the Mg/Fe abundance ratio in chemical evolution
 studies with quasar emission lines, depends largely on the cloud gas
 density. By correcting this density dependence, we propose new
 diagnostics of the Mg/Fe abundance ratio for a broad line region
 cloud. Comparing the derived Mg/Fe abundance ratios with chemical
 evolution models, we suggest that $\alpha$-enrichment by mass loss from
 metal-poor intermediate-mass stars occurred at $z\sim2$ or earlier.
\end{abstract}

\keywords{galaxies: active --- quasars: emission lines --- galaxies:
abundances}

\section{Introduction}
Investigating heavy-element abundances of distant objects over a range
of redshift is essential for understanding the history of star formation
in the Universe.  In particular, the abundance ratio of
$\alpha$-elements such as O, Ne, Mg, and so on relative to Fe has drawn
intense research interest. As is commonly known from nucleosynthesis
calculations, type II supernovae (SNe II) from massive stars mainly
supply $\alpha$-elements, while type Ia supernovae (SNe Ia) from binary
systems mainly supply iron (e.g.,
\citealt{1997NuPhA.616...79N,1997NuPhA.621..467N}).  Since the lifetime
of SN Ia is estimated to be typically $t_\mathrm{Ia} \sim 1$
Gyr\footnote{Recent studies, however, show that the $t_\mathrm{Ia}$
distribution is more weighted toward a shorter life time for SNe Ia
(e.g., \citealt{2008PASJ...60.1327T}; \citealt{2012MNRAS.426.3282M}).}
(\citealt{1983A&A...118..217G}; \citealt{1986A&A...154..279M}) which is
1--2 orders longer than that of SN II, the iron enrichment should be
delayed behind $\alpha$, causing a break in
[$\alpha$/Fe]\footnote{$[\alpha/\mathrm{Fe}] \equiv
\log(n_\alpha/n_\mathrm{Fe}) - \log(n_\alpha/n_\mathrm{Fe})_{\odot}$,
where $n_x$ represents the number density of the element $x (=\alpha$ or
Fe).}  at the elapsed time $t\sim t_\mathrm{Ia}$ from the formation of
the first stars (e.g.,
\citealt{1993ApJ...418...11H,1999ARA&A..37..487H};
\citealt{1996ApJ...462..266Y,1998ApJ...507L.113Y}).  Such a break,
although confirmed from observations of long-lived metal-poor stars in
the solar neighborhood, still requires confirmation at high redshift
corresponding to $t\sim t_\mathrm{Ia}$ in the early Universe.

In this context, quasars offer a valuable tool for exploring the
$[\alpha$/Fe] break at high redshift, because (i) their spectra show
plenty of emission and absorption lines of metals including $\alpha$
elements and iron, and (ii) their brightness is great enough to provide
spectra of good quality even from a far distance at $z > 7$ (e.g.,
\citealt{2011Natur.474..616M}). Measuring the emission-line flux ratio
is one method that can be used to derive the abundance ratio, and the
emission lines from \ion{Mg}{2} and \ion{Fe}{2} have been considered as
an ideal line pair. Their similar ionization energies would cause these
two ions to coexist at a similar location in a broad-line region (BLR)
cloud.  This assumption is partly supported by recent reverberation
mapping studies, which imply that the \ion{Fe}{2} emission region in
quasars exists at a similar location to H$\beta$ that represents other
broad emission lines including \ion{Mg}{2} (e.g.,
\citealt{2014ApJ...783L..34C}, \citealt{2016ApJ...818...30S}).\footnote{
These reverberation studies, however, do not exclude the possibility
that the \ion{Fe}{2} emission region could be twice as large as that of
H$\beta$, so the locations of the \ion{Fe}{2} and \ion{Mg}{2} regions
may not be the same (\citealt{2013ApJ...769..128B};
\citealt{2014ApJ...783L..34C}).}  Therefore, the flux ratio would be at
least linearly related to the abundance ratio (e.g.,
\citealt{1999ARA&A..37..487H}).

\ion{Fe}{2} emission lines observed in quasar spectra were thus eagerly
studied in the 1970s and 1980s (e.g., \citealt{1977ApJ...215..733O};
\citealt{1978ApJS...38..187P,1978ApJ...226..736P};
\citealt{1979A&A....72..293C,1980A&A....83..190C};
\citealt{1981ApJ...250..478K}; \citealt{1981A&A...102..321J};
\citealt{1981ApJ...251..451G}). From their photoionization calculations
and comparison with observations, \cite{1983ApJ...275..445N} and
\cite{1985ApJ...288...94W} showed that iron is presumably overabundant
in comparison with the solar value for quasars, having large \ion{Fe}{2}
flux relative to \ion{Mg}{2}.  Under these circumstances, the
\ion{Mg}{2}/\ion{Fe}{2} flux ratio was assumed to be a first-order proxy
for the Mg/Fe abundance ratio, and has been measured over a wide range
of redshift extending to $z\sim7$ (e.g., \citealt{1996ApJ...470L..85K};
\citealt{1999ApJ...515..487T}; \citealt{2003ApJ...587L..67F};
\citealt{2003ApJ...594L..95B}; \citealt{2003ApJ...596L.155M};
\citealt{2002ApJ...564..581D,2003ApJ...596..817D};
\citealt{2002ApJ...565...63I,2004ApJ...614...69I};
\citealt{2006ApJ...650...57T}; \citealt{2007AJ....134.1150J};
\citealt{2007ApJ...669...32K}; \citealt{2009MNRAS.395.1087S};
\citealt{2011ApJ...739...56D,2014ApJ...790..145D}; and references
therein).  However, the measured flux ratios show a large scatter beyond
measurement errors, especially at high redshift, preventing us from
finding any signature for the [Mg/Fe] break.

Doubt is then cast on the usual assumption that the
\ion{Mg}{2}/\ion{Fe}{2} flux ratio is a first-order proxy for the Mg/Fe
abundance ratio
(\citealt{2011ApJ...739...56D}). \cite{2004ApJ...615..610B}
systematically investigated \ion{Fe}{2} emission lines by using the
photoionization {\sc Cloudy} code combined with the 371-level Fe$^+$
model (\citealt{1999ApJS..120..101V}). Their calculations show that the
strength of the \ion{Fe}{2} emission lines depends not only on iron
abundance, but also on gas density, the ionization parameter, column
density, and microturbulence.  \cite{2003ApJ...592L..59V} also argued
the importance of such dependence using their original 830-level Fe$^+$
model. Therefore, the large scatter observed for \ion{Mg}{2}/\ion{Fe}{2}
may be accounted for by the effects of non-abundance parameters that may
differ from quasar to quasar.

In fact, \cite{2011ApJ...736...86D} found that \ion{Mg}{2}/\ion{Fe}{2}
strongly correlates with the Eddington ratio. They also argued that in
most plausible scenarios connecting active galactic nuclei (AGNs) and
starburst activities (e.g., \citealt{2007ApJ...671.1388D}), the delay
between the two events is less than 1 Gyr, so $\alpha$-elements should
be enhanced relative to iron during the active phase of AGNs. An
anticorrelation between \ion{Mg}{2}/\ion{Fe}{2} and the Eddington ratio
is then naturally expected, whereas the observation contradicts it. These
results imply the existence of non-abundance parameters that would
seriously affect \ion{Mg}{2}/\ion{Fe}{2}. In summary, there is growing
evidence that the \ion{Mg}{2}/\ion{Fe}{2} flux ratio is only a
``second-order'' proxy for the Mg/Fe abundance ratio. However, there
have been few reports as to whether the abundance information can be
extracted quantitatively from observables against these non-abundance
effects.

The purpose of this paper is to invent reliable diagnostics of the Mg/Fe
abundance ratio for a BLR cloud. A sufficient number of quasars selected
from the Sloan Digital Sky Survey (SDSS; \citealt{2000AJ....120.1579Y})
are analyzed to measure their \ion{Mg}{2}/\ion{Fe}{2} flux ratios, and
photoionization calculations are carried out to interpret
\ion{Mg}{2}/\ion{Fe}{2} in terms of the Mg/Fe abundance ratio.  This
paper is organized as follows. In Section 2, we describe the sample
selection and the data reduction of SDSS quasars.  In Section 3,
measured emission-line strengths of the SDSS quasars are summarized. In
Section 4, parameter setting for the photoionization model is described,
and the results are given in Section 5. In Section 6, we discuss the
dependence of \ion{Mg}{2}/\ion{Fe}{2} on non-abundance parameters,
propose new abundance diagnostics for a BLR cloud, and compare the
derived Mg/Fe abundance ratios with chemical evolution models to
constrain star formation history. In Section 7, a summary and conclusion
are given.  Throughout this paper, we assume $\Lambda$CDM cosmology,
with $\Omega_\Lambda=0.7$, $\Omega_M=0.3$, and $H_0=70\ \mathrm{km\
s^{-1}\ Mpc^{-1}}$.

\section{SDSS Data Analysis}
Our sample is selected from the quasar catalog of the SDSS Data Release
7 (DR7), in which 105,783 spectroscopically confirmed quasars are
included (\citealt{2010AJ....139.2360S}). The following two criteria are
adopted for our sample selection. Firstly, a rest frame wavelength range
of 2200--3500\AA\ is available; this is required to measure both
\ion{Fe}{2} and \ion{Mg}{2} emission lines, and the continuum level
around them. Since the wavelength range covered by the spectroscopic
observation in the SDSS DR7 is 3800--9200\AA, this criterion confines
the redshift range within $0.72 < z < 1.63$.  Secondly, the median value
of the signal to noise ratios (S/Ns) in the entire wavelength range of
the SDSS spectra is larger than 10 pix$^{-1}$; from experience, this is
required for accurate measurement of \ion{Fe}{2} and \ion{Mg}{2}
emission lines. We find that 17,468 objects fulfill these two
criteria. According to the catalog given by \cite{2011ApJS..194...45S},
254 objects out of 17,468 are flagged as broad absorption line (BAL)
quasars. Although these BAL quasars are not excluded in the following
analysis, we note that whether or not these are excluded from our sample
affects none of the conclusions of this paper.

The fluxes of the \ion{Mg}{2} and \ion{Fe}{2} emission lines are
measured for the selected quasars in almost the same way as that used by
\cite{2011ApJ...739...56D}. Firstly, the following continuum model is
fitted to each spectrum:
\begin{equation}
 F_\lambda = F_\lambda^{\mathrm{PL}}(\alpha,\beta) +
  F_\lambda^{\mathrm{BaC}} + F_\lambda^{\mathrm{FeII}}(\gamma),
\end{equation}
where $F_\lambda^{\mathrm{PL}}$ is a power-law continuum flux emitted
from an accretion disk, $F_\lambda^{\mathrm{BaC}}$ is a Balmer continuum
flux and $F_\lambda^{\mathrm{FeII}}$ is a \ion{Fe}{2} pseudo-continuum
flux. Following \cite{2011ApJ...739...56D}, we adopt the Balmer
continuum model given by \cite{1982ApJ...255...25G} for
$F_\lambda^{\mathrm{BaC}}$; the shape and the flux ratio against the
power-law component are fixed to those adopted by
\cite{2011ApJ...739...56D}. Unlike \cite{2011ApJ...739...56D}, however,
we use the \ion{Fe}{2} template given by \cite{2006ApJ...650...57T}
instead of that of \cite{2001ApJS..134....1V}; this is because the
latter does not cover the wavelength range around the \ion{Mg}{2}
emission line.  Prior to fitting, the \ion{Fe}{2} template is broadened
by convolution with a Gaussian function for which the full width at half
maximum (FWHM) is fixed\footnote{Note that, as
\cite{2011ApJ...739...56D} pointed out, the measured \ion{Fe}{2} flux
depends very little on the FWHM adopted for the convolved Gaussian
function.}  at 2,000 km s$^{-1}$. There are thus three free parameters:
the power-law slope ($\alpha$), the normalization of the power-law
continuum flux ($\beta$), and the normalization of the \ion{Fe}{2}
pseudo-continuum flux ($\gamma$). The best fit parameters are obtained
by performing $\chi^2$ minimization with the IDL procedure {\tt
MPFIT.pro} (\citealt{2009ASPC..411..251M}). \ion{Fe}{2} flux is
calculated by integrating the fitted \ion{Fe}{2} template in a
wavelength range of 2200--3090\AA. Then, fitting of the \ion{Mg}{2}
$\lambda2798$ emission line is performed; two Gaussians are fitted to
the continuum-subtracted spectrum at the rest frame wavelength range of
2700--2900\AA\ with the {\tt MPFIT.pro} procedure. Both the flux and the
FWHM of \ion{Mg}{2} are calculated from the sum of the two fitted
Gaussians.  In these processes, there are failures in the fittings for
36 quasars.  However, since the number is too small to affect the
result, we have decided to exclude them from the following analysis.

The mass of a black hole (BH), or $M_{\mathrm{BH}}$, is estimated from
the \ion{Mg}{2} FWHM and the continuum luminosity at 3000\AA\ by using
the virial mass estimate formula (\citealt{2009ApJ...699..800V}):
\begin{eqnarray}
\log \left( \frac{M_{\mathrm{BH}}}{M_{\odot}} \right) = 6.86 + 2 \log
 \left( \frac{\mathrm{FWHM(MgII)}}{1,000\ \mathrm{km\ s^{-1}}} \right)
 \nonumber \\
 + 0.5 \log \left( \frac{\lambda L_\lambda(3000\mathrm{\AA})}{10^{44}\
	     \mathrm{erg\ s^{-1}}} \right). \label{eq:bhmass}
\end{eqnarray}
The Eddington luminosity is defined as the luminosity at which the
radiation force acting on an electron--proton pair is balanced with the
gravitation force exerted on the pair. For a BLR cloud orbiting the
central BH, the Eddington luminosity is given by
\begin{equation}
 L_{\mathrm{Edd}} = \frac{4\pi Gc m_p}{\sigma_e}M_{\mathrm{BH}}, \label{eq:edd_lumi}
\end{equation}
where $G$ is the gravitational constant, $c$ is the speed of light,
$m_p$ is the proton mass, and $\sigma_e$ is the Thomson scattering cross
section (\citealt{1997iagn.book.....P}). Following
\cite{2011ApJS..194...45S}, we estimate the bolometric luminosity from
the measured monochromatic luminosity at 3000\AA\ with the bolometric
correction formula:
\begin{equation}
 L_{\mathrm{bol}} = 5.15 \lambda L_\lambda(3000\mathrm{\AA}). \label{eq:bol_lumi}
\end{equation}
From equations (\ref{eq:bhmass})--(\ref{eq:bol_lumi}), the Eddington ratio
is written as
\begin{eqnarray}
 \log \left( \frac{L_{\mathrm{bol}}}{L_{\mathrm{Edd}}} \right) = 
  -0.249 - 2 \log \left( \frac{\mathrm{FWHM(MgII)}}{1,000\ \mathrm{km\
		s^{-1}}} \right) \nonumber \\
 + 0.5 \log \left( \frac{\lambda L_\lambda(3000\mathrm{\AA})}{10^{44}\ \mathrm{erg\ s^{-1}}} \right).
\end{eqnarray}
For all the quasars in our sample, the Eddington ratio is evaluated 
using this formula.

\begin{figure*}[t]
  \epsscale{1.0} \plotone{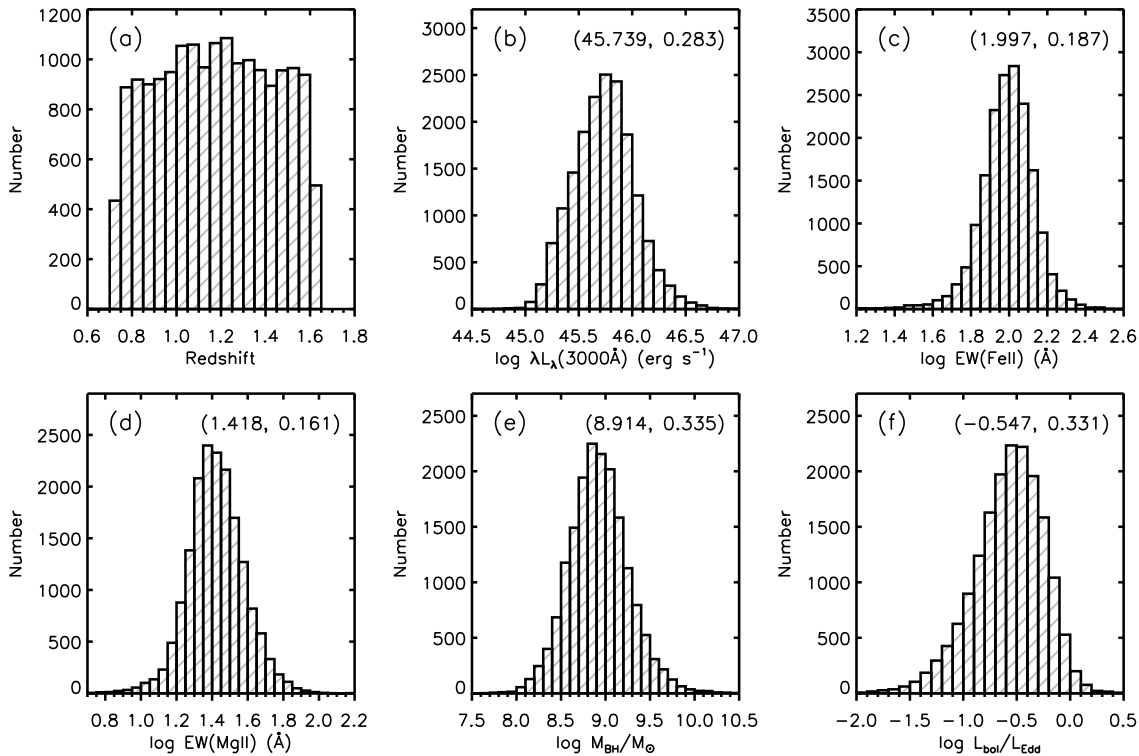}
  \caption{Histograms of the measured values for our sample of SDSS quasars: (a)
  redshift, (b) monochromatic luminosity at 3000\AA, (c) EW(\ion{Fe}{2}), (d)
  EW(\ion{Mg}{2}), (e) BH mass, and (f) Eddington ratio. On each panel except
  for (a), both the median value and the standard deviation are
  given in the upper-right corner in parentheses (median, 1$\sigma$).}
  \label{fig:measure}
\end{figure*}

\begin{figure}[t]
 \epsscale{1.0} \plotone{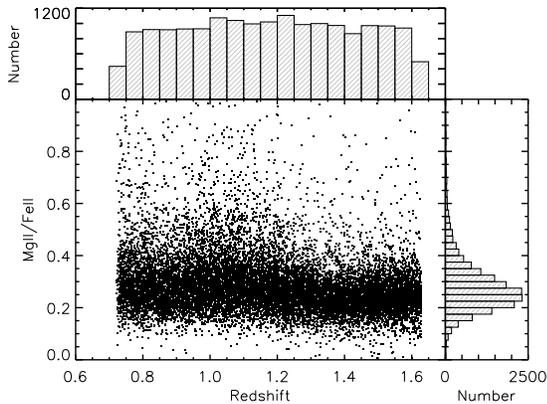}
 \caption{\ion{Mg}{2}/\ion{Fe}{2} flux ratio versus redshift for our sample 
 of SDSS quasars. The histograms of flux ratio and redshift are shown
 in the right and upper panels, respectively.}
 \label{fig:mgiifeii_redshift}
\end{figure}

In summary, our final sample consists of $17,432$ quasars in the
redshift range of $0.72 < z < 1.63$.  Figure \ref{fig:measure} shows
histograms of the measured values of various quantities such as
redshift, monochromatic luminosity at 3000\AA, rest-frame equivalent
widths (EWs) of \ion{Fe}{2} and \ion{Mg}{2}, BH mass, and the Eddington
ratio of our sample of SDSS quasars.  The EWs of \ion{Fe}{2} and
\ion{Mg}{2} emission lines are calculated by dividing the measured flux
by the continuum flux density at 3000\AA. As previous studies have
pointed out (\citealt{2003ApJS..145..199M};
\citealt{2011ApJ...736...86D}), {\tt MPFIT.pro} likely underestimates
measurement errors, because it does not account for potential systematic
errors related to, e.g., \ion{Fe}{2} pseudocontinuum
subtraction. Therefore, typical measurement errors are estimated by
Monte Carlo simulations (see \citealt{2011MNRAS.410.1018S}) as follows:
16\% for EW(\ion{Fe}{2}), 7.2\% for EW(\ion{Mg}{2}), 7.9\% for
FWHM(\ion{Mg}{2}), and 10\% for $\lambda L_\lambda(3000\mathrm{\AA})$.
Figure \ref{fig:mgiifeii_redshift} plots the measured
\ion{Mg}{2}/\ion{Fe}{2} flux ratios against their redshifts. It is clear
that \ion{Mg}{2}/\ion{Fe}{2} shows no significant evolution at $z\sim
1$--$2$. This result is consistent with \cite{2002ApJ...565...63I}, in
whose analysis an earlier version of the SDSS data was used.

\section{EW(\ion{Fe}{2}) versus EW(\ion{Mg}{2}) diagram}
\begin{figure*}[t]
  \epsscale{1.0} \plotone{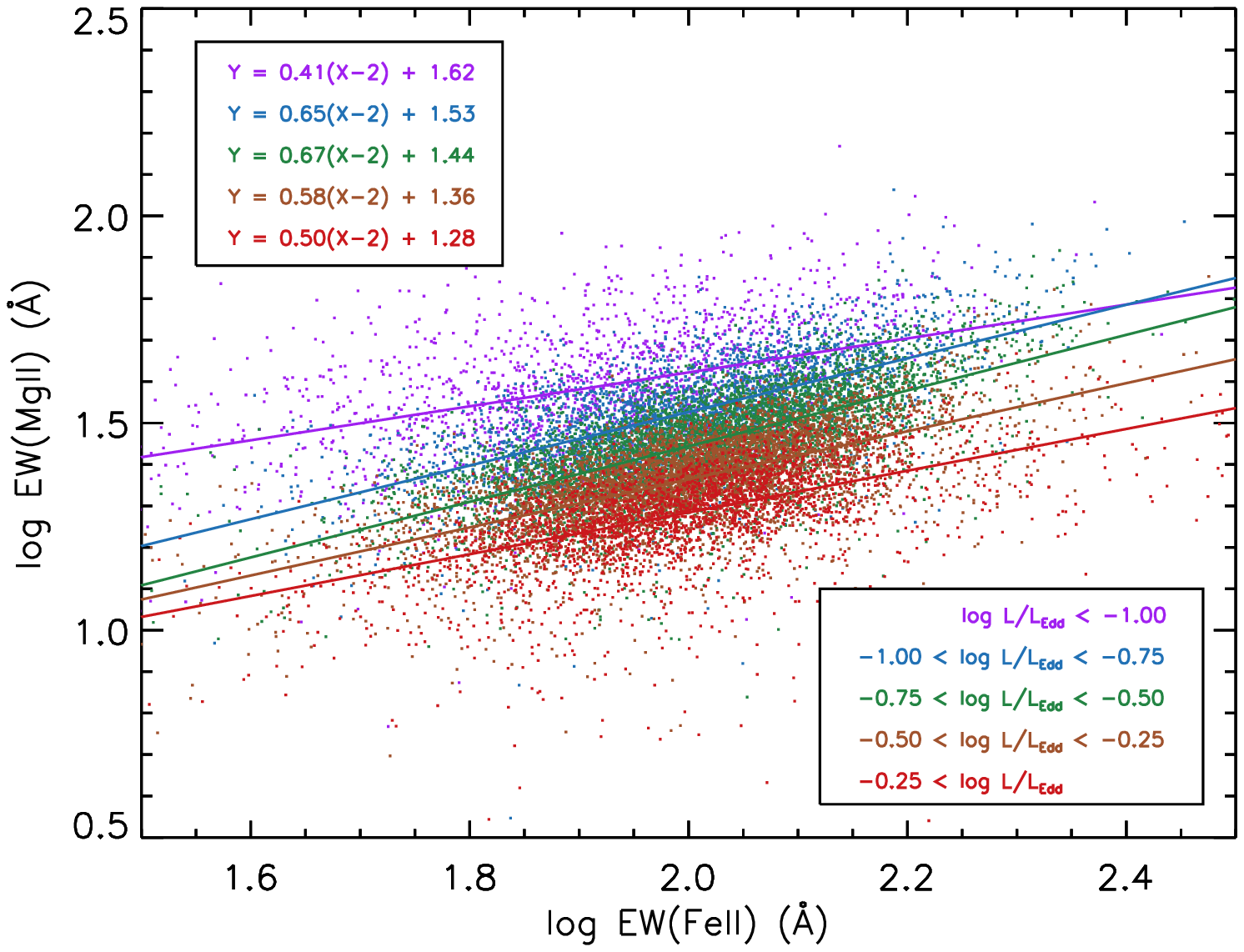}
  \caption{EW(\ion{Mg}{2}) versus EW(\ion{Fe}{2}) diagram. The SDSS
  quasars from our sample in different ranges of the Eddington ratio are
  colored accordingly: $L_{\mathrm{bol}}/L_{\mathrm{Edd}} < 10^{-1.0}$
  (purple), $10^{-1.0}$--$10^{-0.75}$ (blue), $10^{-0.75}$--$10^{-0.5}$
  (green), $10^{-0.5}$--$10^{-0.25}$ (brown), and $> 10^{-0.25}$
  (red). Linear regression analyses are performed for these sub-samples,
  and the results are given in the inset in the upper-left corner.}
  \label{fig:ew_mgii_feii}
\end{figure*}

\begin{figure*}[t]
  \epsscale{1.0} \plotone{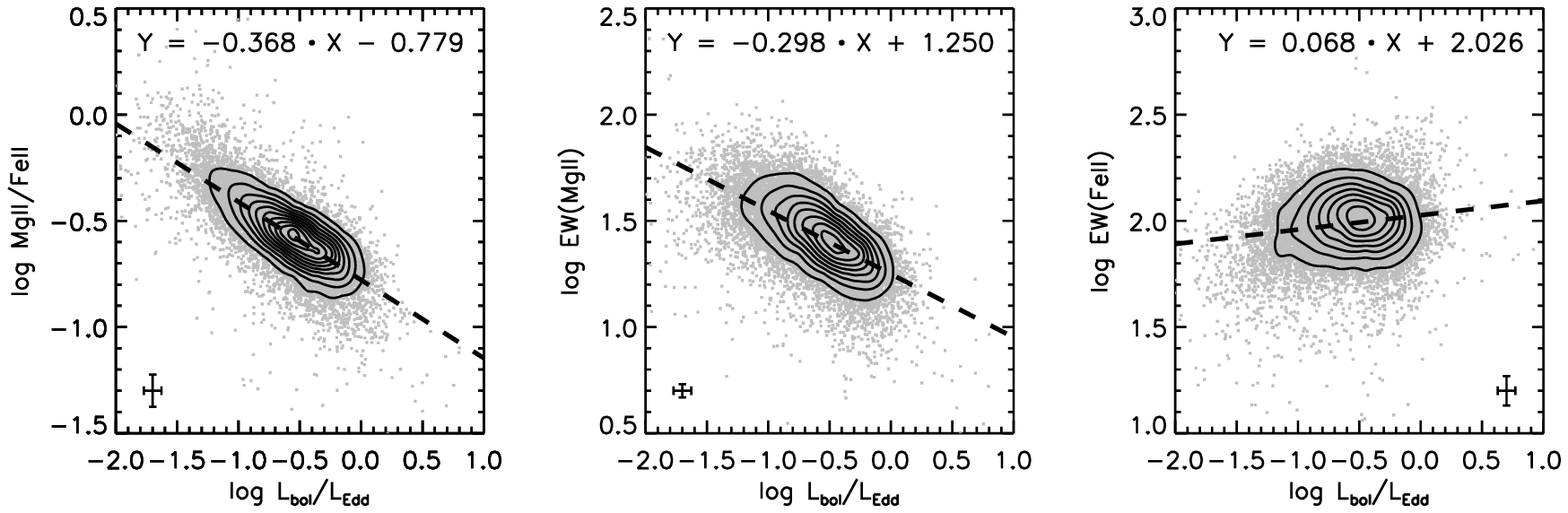}
  \caption{Dependence of emission-line strength on the Eddington ratio
  for our sample of SDSS quasars. (a) \ion{Mg}{2}/\ion{Fe}{2}, (b)
  EW(\ion{Mg}{2}), (c) EW(\ion{Fe}{2}). The contours indicate local data
  point densities with a grid size of $\Delta=0.01$ dex. The dashed line
  indicates a linear regression line. The typical size of measurement
  errors are indicated at the corner of each panel.}
  \label{fig:fluxratio_ew_edd}
\end{figure*}

From comparison of the measured values, we find that EW(\ion{Mg}{2})
positively correlates with EW(\ion{Fe}{2}), as seen from the $\log$
EW(\ion{Fe}{2}) versus $\log$ EW(\ion{Mg}{2}) plot in Figure
\ref{fig:ew_mgii_feii}.  To examine how this correlation depends on the
Eddington ratio, we divide our sample into five sub-samples: (i)
$L_{\mathrm{bol}}/L_{\mathrm{Edd}} < 10^{-1.0}$, (ii)
$10^{-1.0}$--$10^{-0.75}$, (iii) $10^{-0.75}$--$10^{-0.5}$, (iv)
$10^{-0.5}$--$10^{-0.25}$, and (v) $> 10^{-0.25}$. These sub-samples are
shown by different colors in Figure \ref{fig:ew_mgii_feii}. It is
evident that quasars with larger $L_{\mathrm{bol}}/L_{\mathrm{Edd}}$
tend to have smaller EW(\ion{Mg}{2}).

Dependence of \ion{Mg}{2}/\ion{Fe}{2}, \ion{Mg}{2},\ and \ion{Fe}{2}\ on
$L_{\mathrm{bol}}/L_{\mathrm{Edd}}$ is highlighted in Figure
\ref{fig:fluxratio_ew_edd}.  Negative correlation against
$L_{\mathrm{bol}}/L_{\mathrm{Edd}}$ is clearly seen for
\ion{Mg}{2}/\ion{Fe}{2}\ and EW(\ion{Mg}{2}). On the other hand,
EW(\ion{Fe}{2}) has a tentative, positive correlation with
$L_{\mathrm{bol}}/L_{\mathrm{Edd}}$. We have performed linear regression
analysis using the BCES(Y\verb+|+X) method
(\citealt{1996ApJ...470..706A}), which takes into account measurement errors
on both variables. The result is summarized in Table
\ref{tab:correlation}, where the calculated Spearman's correlation
coefficients are also listed.

It is worth mentioning the preceding research by
\cite{2011ApJ...736...86D}.  They analyzed 4,178 Seyfert 1 galaxies and
quasars selected from SDSS DR4 for which redshifts are $z \le 0.8$.
They reported that the \ion{Fe}{2}/\ion{Mg}{2} flux ratio positively
correlates with the Eddington ratio, which is consistent with our
result.  Note that they analyzed both UV and optical \ion{Fe}{2}
emission lines, therefore the redshift range of their sample is
restricted.  Compared with their analysis, we have used the later
version of the SDSS data and concentrated on the UV \ion{Fe}{2} emission
line, which has increased the sample size by almost four times and
widened the redshift range.  One of the most interesting results of
\cite{2011ApJ...736...86D} is that narrow \ion{Fe}{2} emission in
optical has a stronger correlation with the Eddington ratio than
\ion{Mg}{2}.  Whether the narrow \ion{Fe}{2} emission line exists in UV
is unclear and investigating it is beyond the scope of this paper, but
we may safely say that the slight correlation between the UV \ion{Fe}{2}
and the Eddington ratio seen in Figure \ref{fig:fluxratio_ew_edd}
implies that contribution from the narrow \ion{Fe}{2} emission in UV is
small.

The \mgii/\feii--$L_{\mathrm{bol}}/L_{\mathrm{Edd}}$ correlation is thus
evident from observation, but the background physics behind it has not
been well understood so far. In the following part of this paper, we
investigate the origin of such emission-line correlations with the
Eddington ratio.

\begin{deluxetable}{lrrr}
\tablecaption{Correlations with the Eddington ratio.\label{tab:correlation}}
\tablewidth{0pt}
\tablehead{
 \colhead{Quantity (y)} & \colhead{Slope (a)} & \colhead{Intercept (b)} & \colhead{$\rho^\dagger$}
}
\startdata
\ion{Mg}{2}/\ion{Fe}{2} & $-0.368\pm0.005$ & $-0.779\pm0.003$ & $-0.73$ \\
EW \ion{Mg}{2} (\AA)    & $-0.298\pm0.005$ & $+1.250\pm0.003$ & $-0.64$ \\
EW \ion{Fe}{2} (\AA)    & $+0.068\pm0.007$ & $+2.026\pm0.004$ & $+0.08$
\enddata
\tablecomments{Linear regression analysis is performed with 
$\log y = a\times \log L_\mathrm{bol}/L_\mathrm{Edd} + b$.}
\tablenotetext{$\dagger$}{Spearman's rank correlation coefficient.}

\end{deluxetable}

\section{Photoionization Calculations}

Photoionization calculations have been carried out to investigate how
\ion{Mg}{2} and \ion{Fe}{2} emission lines emitted from a BLR cloud
depend on non-abundance parameters such as gas density, the ionization
parameter, total H column density, spectral energy distribution (SED) of
the illuminating source, and microturbulence. The values of these
parameters are varied in their plausible respective ranges in
calculations. In this section, we first explain the constraints on these
parameters inferred from recent studies of BLR reverberation
mapping. Then, we explain our choice of fiducial parameter values for a
BLR cloud.

\subsection{Constraint on the property of a BLR cloud inferred from reverberation
  mapping studies} \label{sec:RM}

Recent studies of BLR reverberation mapping (e.g.,
\citealt{2004ApJ...613..682P}; \citealt{2005ApJ...629...61K};
\citealt{2006ApJ...644..133B,2009ApJ...697..160B,2013ApJ...767..149B})
have revealed the so-called radius--luminosity relationship; the
reverberation radius is larger for more luminous AGNs. From naive 
theoretical consideration, $R \propto L^{0.5}$ is naturally expected
under the assumption that the central radiation field is the only source
of heating and ionization. In fact, \cite{2013ApJ...767..149B} measured
H$\beta$ lag times of 41 nearby AGNs from BLR reverberation mapping
observations, and showed that the radius--luminosity relationship holds
for AGNs over four orders of magnitude in luminosity.

Although there exist few studies of reverberation mapping for the
\ion{Mg}{2} emitting region, \cite{2006ApJ...647..901M} report that the
\ion{Mg}{2} reverberation radius is almost the same as H$\beta$ for NGC
4151. Recently, \cite{2016ApJ...818...30S} report \ion{Mg}{2} lag
detection of 6 quasars at $0.3 < z < 0.8$, which also supports overlap
between the regions in which \mgii\ and H$\beta$ originate. Under the
situation that no reliable lag detection at $z > 0.8$ is available, it
is reasonable to assume that the \ion{Mg}{2} emission line is created in
the same region as H$\beta$ even at $z > 0.8$.  Thus, we assume that the
radius--luminosity relationship for \ion{Mg}{2} is the same as for
H$\beta$ given in \cite{2013ApJ...767..149B}, and is written as
\begin{equation}
\frac{R_{\mathrm{MgII}}}{10\ \text{lt-days}} = (3.40 \pm 0.20) \left[
						       \frac{\lambda
						       L_\lambda(5100\mathrm{\AA})}{10^{44}\ 
						       \mathrm{erg\ s^{-1}}}
						       \right]^{0.5}, \label{eq:sizelumi}
\end{equation}
where $R_{\mathrm{MgII}}$ is the \ion{Mg}{2} reverberation radius. Note
that we here adopt the power index of 0.5, as theoretically
expected. The coefficient in equation (\ref{eq:sizelumi}) is determined
by fitting a slope-fixed straight line to the data of
\cite{2013ApJ...767..149B}.

This radius--luminosity relationship places an important constraint on
the physical quantities of a BLR cloud.  Let $U$ be the ionization
parameter defined as
\begin{equation}
 U = \frac{\Phi(\mathrm{H})}{n_{\mathrm{H}} c}, \label{eq:ion_param}
\end{equation}
where $n_\mathrm{H}$ is the hydrogen number density of the gas, $c$ is
the speed of light, and $\Phi(\mathrm{H})$ is the incident
hydrogen-ionizing photon flux. Furthermore, $\Phi(\mathrm{H})$ can be
reduced to
\begin{equation}
 \Phi(\mathrm{H}) = \int_{\nu_0}^\infty \frac{F_\nu}{h\nu} d\nu =
  \frac{1}{4\pi R^2} \int_{\nu_0}^\infty \frac{L_\nu}{h\nu} d\nu \equiv
  \frac{Q(\mathrm{H})}{4\pi R^2}, \label{eq:ion_flux}
\end{equation}
where $\nu_0$ is the frequency corresponding to the ionization energy of
H$^0$ (13.6 eV), $F_\nu$ is the flux density of the incident continuum
at the illuminated face of a cloud, $L_\nu$ is the luminosity of the
central source at frequency $\nu$, and $Q(\mathrm{H})$ is the number of
hydrogen-ionizing photons emitted per second from the source. From
equations (\ref{eq:sizelumi})--(\ref{eq:ion_flux}), the product of
$n_\mathrm{H}$ and $U$ for the \ion{Mg}{2} emitting region yields
\begin{equation}
 n_{\mathrm{H}}U = (3.42 \pm 0.40) \times 10^{-2} \left( \frac{Q(\mathrm{H})}{\lambda
  L_\lambda(5100\mathrm{\AA})} \right) \ \mathrm{cm^{-3}}, \label{eq:nU}
\end{equation}
where the term $Q(\mathrm{H})/\lambda L_\lambda(5100\mathrm{\AA})$
depends only on the SED of the incident continuum.

Following the user manual of the {\sc Cloudy} code (last described by
\citealt{2013RMxAA..49..137F}), we have modeled the SED of an AGN as
follows:
\begin{equation}
 F_\nu = \nu^{\alpha_\mathrm{uv}} \exp(-h\nu/kT_{\mathrm{BB}})
  \exp(-kT_{\mathrm{IR}}/h\nu) + a \nu^{\alpha_{\mathrm{x}}}, \label{eq:AGN_SED}
\end{equation}
where the first term on the right-hand side indicates the so-called {\it
big blue bump} component, and the second term indicates the non-thermal
X-ray component.  The former is modeled as a power law with the index
$\alpha_{\mathrm{uv}}$, which has exponential cutoffs at the energies of
$kT_{\mathrm{BB}}$ and $kT_{\mathrm{IR}}$; the latter is modeled as a
power law with the index $\alpha_{\mathrm{x}}$, which is truncated at
energies less than 1.36 eV in order to prevent it from extending into
the infrared part, and falls off as $\nu^{-2}$ at energies larger than
100 keV.  The coefficient $a$, which determines a fraction between the
two components, is uniquely determined if the well-used optical to X-ray
spectral index $\alpha_{\mathrm{ox}}$\footnote{$\alpha_{\mathrm{ox}}
\equiv \log \left[ f_\nu(2\ \mathrm{keV})/f_\nu(2500\mathrm{\AA})
\right]/\log(403.3)$.}  and the spectral indices $\alpha_{\mathrm{uv}},
\alpha_{\mathrm{x}}$ are once specified.  Following
\cite{1997ApJS..108..401K}, we adopt $\alpha_{\mathrm{ox}}=-1.40,\
\alpha_{\mathrm{uv}}=-0.50,\ \alpha_{\mathrm{x}}=-1.0,\
T_\mathrm{BB}=10^6\ \mathrm{K}$, and $kT_\mathrm{IR}=0.136$ eV to
reproduce the SED of a typical AGN.  This results in $a=3.5\times10^6$
and $Q(\mathrm{H})/\lambda L_\lambda(5100\mathrm{\AA})=8.36 \times
10^{10}\ \mathrm{erg^{-1}}$.  Substituting the latter value into
equation (\ref{eq:nU}), we obtain $n_\mathrm{H}U = (2.86 \pm 0.33)
\times 10^9\ \mathrm{cm^{-3}}$ or $n_\mathrm{H}U \sim 10^{9.5}\
\mathrm{cm^{-3}}$. To evaluate the dependence of SED diversity observed
in quasars (e.g., \citealt{2006ApJS..166..470R}), we have varied the
X-ray power-law index $\alpha_{\mathrm{x}}$ and the spectral index
$\alpha_{\mathrm{ox}}$ in a wide range of $-2 \le \alpha_{\mathrm{x}}
\le 0$ and $-2 \le \alpha_{\mathrm{ox}} \le -1$, respectively, from the
above SED.  The resultant value of $n_\mathrm{H}U$, however, changes
only by 0.2 dex.  It is thus appropriate to assume that all quasars in
our sample follow $n_\mathrm{H}U \sim 10^{9.5}\ \mathrm{cm^{-3}}$.  We
note that this constraint is always taken into account in the following
photoionization calculations of the \ion{Mg}{2} and \ion{Fe}{2} emission
lines.

\subsection{Parameter setting}
We use the photoionization simulation code of {\sc Cloudy} version
13.02, combined with the 371 level Fe$^+$ model that includes all energy
levels up to 11.6 eV (\citealt{1999ApJS..120..101V}). The parameters for
a BLR cloud required to run the code are the hydrogen gas density
$n_\mathrm{H}$, the ionization parameter $U$, the SED of the
illuminating source, the elemental abundance in the gas, and the column
density that determines the outer edge of the cloud.

From the measured emission-line strengths and their ratios, the BLR gas
density is estimated to be $n_\mathrm{H}=10^{10}$--$10^{12}\
\mathrm{cm^{-3}}$ (e.g., \citealt{1989ApJ...347..640R};
\citealt{1992ApJ...387...95F}; \citealt{1996ApJ...461..664B}). A
fiducial value of the ionization parameter is estimated to be $U \sim
10^{-1}$ (e.g., \citealt{2014MNRAS.438..604B}), but
\cite{2007ApJ...663..781M} report that an \ion{O}{1} emitting region in
the BLR cloud, from which the \ion{Mg}{2} and \ion{Fe}{2} emission lines
are also expected to originate, has a value of $U \sim
10^{-2.5}$. Combining these implications and the constraint
$n_\mathrm{H} U = 10^{9.5}\ \mathrm{cm^{-3}}$ inferred from the
reverberation mapping studies, we adopt $n_\mathrm{H}=10^{11}\
\mathrm{cm^{-3}}$ and $U=10^{-1.5}$ as the fiducial parameter values.

The SED of the illuminating source is set as in the previous section to
reproduce the SED of a typical AGN, i.e., equation (\ref{eq:AGN_SED})
and the following parameters are adopted: $\alpha_{\mathrm{ox}}=-1.40,\
\alpha_{\mathrm{uv}}=-0.50,\ \alpha_{\mathrm{x}}=-1.0,\
T_\mathrm{BB}=10^6\ \mathrm{K}$, and $T_\mathrm{IR}=0.136$ eV.

Although there are only a few studies of observational constraints on
the column density of the BLR cloud, many photoionization calculations
assume that $N_\mathrm{H}=10^{23}\ \mathrm{cm^{-2}}$ (e.g.,
\citealt{1995ApJ...455L.119B,2004ApJ...615..610B};
\citealt{1997ApJS..108..401K};
\citealt{2008ApJ...673...62M}). \cite{2011MNRAS.410.1018S} proposed that
the optical \ion{Fe}{2} to UV \ion{Fe}{2} flux ratio can be used as a
column density indicator; they found its median value for 884 SDSS
DR5 quasars to be $N_\mathrm{H}=10^{22.8}\ \mathrm{cm^{-2}}$. Accordingly,
we adopt $N_\mathrm{H}=10^{23}\ \mathrm{cm^{-2}}$ as the fiducial value.

\begin{deluxetable}{lrlrlr}
 \tabletypesize{\scriptsize}
 \tablecaption{Solar abundance \label{tab:solar}}
 \tablewidth{0pt}
 \tablehead{
 \colhead{Element} & \colhead{$\log n$} & \colhead{Element} & \colhead{$\log n$}
 & \colhead{Element} & \colhead{$\log n$}
 }
 \startdata
 H  & $0.00$   & Na & $-5.76$ & Sc & $-8.85$ \\
 He & $-1.07$  & Mg & $-4.40$ & Ti & $-7.05$ \\
 Li & $-10.95$ & Al & $-5.55$ & V  & $-8.07$ \\
 Be & $-10.62$ & Si & $-4.49$ & Cr & $-6.36$ \\
 B  & $-9.30$  & P  & $-6.59$ & Mn & $-6.57$ \\
 C  & $-3.57$  & S  & $-4.88$ & Fe & $-4.50$ \\
 N  & $-4.17$  & Cl & $-6.50$ & Co & $-7.01$ \\
 O  & $-3.31$  & Ar & $-5.60$ & Ni & $-5.78$ \\
 F  & $-7.44$  & K  & $-6.97$ & Cu & $-7.81$ \\
 Ne & $-4.07$  & Ca & $-5.66$ & Zn & $-7.44$
 \enddata
 
 \tablecomments{The values indicate the logarithmic number density
 relative to hydrogen.}
\end{deluxetable}

For chemical composition, we assume the solar abundances of
\cite{2009ARA&A..47..481A} as the fiducial, which are summarized in
Table \ref{tab:solar}.  Since Fe and Mg are refractory elements, their
abundances would be largely affected by dust depletion.  However, we
conclude that ignoring the dust depletion in our simulations is
reasonable for the following reason.  \cite{2010ApJ...721.1835S}
proposed that the strength of optical \ion{Fe}{2} emission lines in
AGNs, which may originate from outside the BLR, may largely result from
different degrees of the Fe depletion into dust grains.  On the other
hand, they also proposed that UV \ion{Fe}{2} emission lines, which have
been measured in this paper, originate in the BLR where dust grains
evaporate, and are little affected by the dust depletion.

For metallicity, many studies suggest super-solar
values for BLR clouds. For example, \cite{2002ApJ...564..592H} estimated
the BLR metallicity from emission lines to be 1--3
$Z_{\odot}$. Using similar methods, \cite{2006A&A...447..157N} reported
that the typical metallicity of $2.0 < z < 4.5$ quasars is
$\sim5Z_{\odot}$. Since the redshift of our sample is relatively low
($0.72 < z < 1.63$), we here adopt $Z = 3Z_{\odot}$ as a fiducial
value.

Most of the \ion{Fe}{2} emission lines are emitted from collisionally
excited levels.  However, observed spectra show some \ion{Fe}{2}
emission lines from upper levels which are too high to have been
collisionally excited by thermal electrons. This problem was overcome by
adding microturbulence\footnote{Magnetehydrodynamic waves are proposed
as the origin of microturbulence for BLR clouds
(\citealt{1987MNRAS.228P..47R}; \citealt{2000MNRAS.316..103B}), but it
still remains an open research problem. } to strengthen the photon
pumping effect (e.g., \citealt{1983ApJ...275..445N}). From systematic
photoionization calculations, \cite{2004ApJ...615..610B} showed that
$v_{\mathrm{turb}} \ge 100\ \mathrm{km\ s^{-1}}$ is needed to reproduce
the observed \ion{Fe}{2} spectral feature of a narrow-line Seyfert
galaxy I Zw 1.  In their recent study on \ion{Fe}{2} emission in AGNs,
\cite{2009ApJ...707L..82F} adopted $v_\mathrm{turb}=100\ \mathrm{km\
s^{-1}}$.  Therefore, we adopt $v_\mathrm{turb}=100\ \mathrm{km\
s^{-1}}$ as the fiducial value.

\begin{deluxetable}{ll}
 \tablecaption{Parameters of the baseline model for a BLR cloud \label{tab:baseline}}
 \tablewidth{0pt}
 \tablehead{
 \colhead{Parameter} & \colhead{Fiducial value}
 }
 \startdata
 Hydrogen gas density $n_\mathrm{H}$ & $10^{11}\ \mathrm{cm^{-3}}$ \\
 Ionization parameter $U$ & $10^{-1.5}$ \\
 Column density $N_\mathrm{H}$ & $10^{23}\ \mathrm{cm^{-2}}$ \\
 SED & \cite{1997ApJS..108..401K} \\
 Elemental abundance pattern $x_{i, \odot}$ & Solar
 (Table \ref{tab:solar}) \\
 Metallicity $Z$ & $3Z_{\odot}$ \\
 Microturbulence $v_\mathrm{turb}$ & $100\ \mathrm{km\,s^{-1}}$
 \enddata
\end{deluxetable}
The fiducial parameter values of the baseline model are listed in Table
\ref{tab:baseline}.  To investigate the parameter dependence, each
parameter value is varied while other parameters are fixed in {\sc
Cloudy}.

\section{Results}
\subsection{Ionization structure}

\begin{figure}[t]
  \epsscale{1.0} \plotone{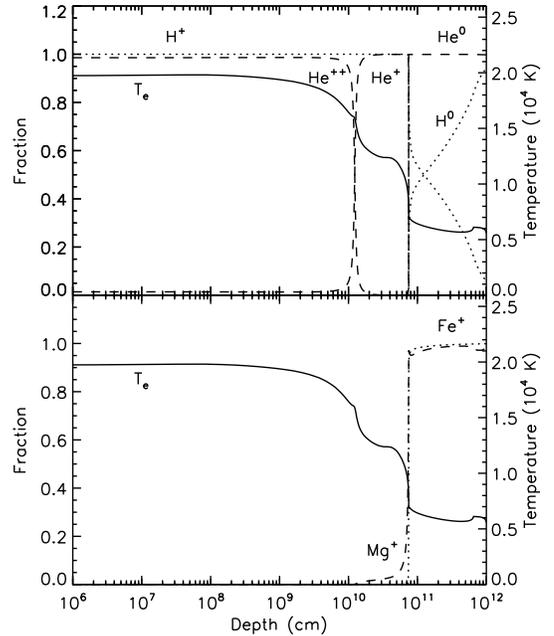}
  \caption{Ionization structure of a photoionized BLR cloud. {\it Top:}
  Fraction of H (dotted lines) and He (dashed lines) ions as a function
  of cloud depth from the illuminated face. The solid line indicates the
  electron temperature. {\it Bottom:} The same as the top panel, but for
  Fe$^+$ (dotted line) and Mg$^+$ (dashed line) ions.}
  \label{fig:ion_fract}
\end{figure}

The calculated ionization structure and electron temperature
distribution in the BLR cloud for the baseline model are shown in Figure
\ref{fig:ion_fract}; the result for H and He is shown in the upper
panel, while that for Mg and Fe is shown in the lower panel. Hydrogen is
fully ionized from the cloud surface up to a depth of $D\sim 10^{11}$
cm, and is partially ionized in the deeper region. Because of the
similarity in the ionization energies of Fe and Mg,\footnote{From
\cite{1996ApJ...465..487V}, the ionization energies of Fe in two
different ion stages are 7.902 eV (Fe$^0$) and 16.19 eV (Fe$^+$), and
those of Mg are 7.646 eV (Mg$^0$) and 15.04 eV (Mg$^+$).} both
\ion{Fe}{2} and \ion{Mg}{2} have generally been assumed to originate
from the partially ionized zone (hereafter PIZ). However, the lower
panel of Figure \ref{fig:ion_fract} indicates that an Mg$^+$ ion can
survive in the fully ionized zone (hereafter FIZ) where almost all iron
is ionized to Fe$^{2+}$ or higher ion stages.

\begin{figure}[t]
  \epsscale{1.0} \plotone{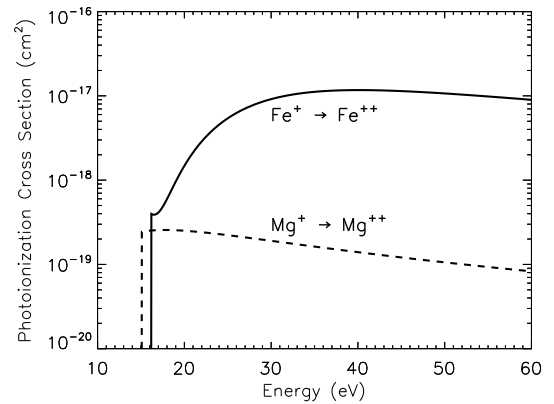}
  \caption{Photoionization cross section as a function of photon energy, 
  taken from \cite{1996ApJ...465..487V}. The solid and dashed lines indicate 
  Fe$^+$ and Mg$^+$, respectively}  \label{fig:photoion_cs}
\end{figure}

The difference between the ionization stages of Mg and Fe in the FIZ is
explained as follows. Since thermal electrons at $T_\mathrm{e} \sim
10^4$ K do not have enough energy to collisionally ionize these ions,
the ionization stage is substantially determined by a balance between
photoionization and recombination. Figure \ref{fig:photoion_cs} shows
the photoionization cross sections of Fe$^+$ and Mg$^+$ as a function of
photon energy, taken from \cite{1996ApJ...465..487V}.  While their
threshold energies of ionization are similar ($\sim15$ eV), the
photoionization cross section of Fe$^+$ is 1--2 orders of magnitude
larger than that of Mg$^+$.  Therefore, the ionization stage of Fe is
higher than that of Mg at the same cloud depth in the FIZ.

 \subsection{Line emissivity}
\begin{figure}[t]
  \epsscale{1.0}
  \plotone{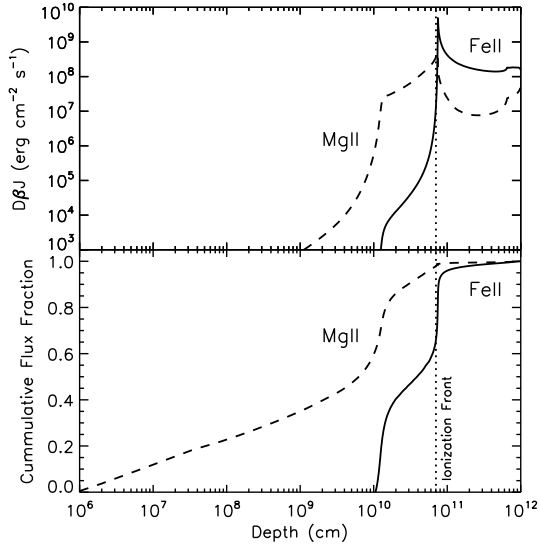} \caption{{\it
  Top:} Line emissivity as a function of cloud depth $D$ for \ion{Fe}{2}
  (solid line) and \ion{Mg}{2} (dashed line). The vertical axis
  indicates $D\beta J$, where $\beta$ is an escape probability of the
  photon, and $J$ is an emissivity.  Note that an emergent emission-line
  flux can be calculated by $\int D\beta J d(\ln D)$. The vertical
  dotted line indicates a boundary between the FIZ on the left and the
  PIZ on the right. {\it Bottom:} The same as the top panel, but the
  vertical axis indicates a cumulative flux fraction.}
  \label{fig:emissivity}
\end{figure}

The calculated emissivities of the \ion{Mg}{2} and \ion{Fe}{2} lines for
the baseline model are shown as a function of cloud depth in the upper
panel of Figure \ref{fig:emissivity}.  Note that the vertical axis is
the line emissivity multiplied by the escape fraction $\beta$ and the
cloud depth $D$.  As expected, the \ion{Fe}{2} flux originates in the
PIZ where the lines are mostly created. On the other hand, the
emissivity of the \ion{Mg}{2} line peaks in the FIZ, meaning that the
\ion{Mg}{2} flux originates in the FIZ, but {\it not} in the PIZ where
Mg predominantly exists in the form of Mg$^+$ ions. This is more clearly
shown in the lower panel of Figure \ref{fig:emissivity}, where the
vertical axis indicates the cumulative flux fraction.

\begin{figure}[t]
  \epsscale{1.0} \plotone{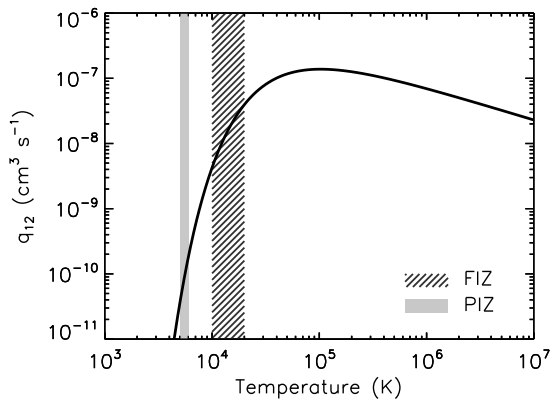}
 \caption{\ion{Mg}{2} collisional excitation rate as a function of
 electron temperature. The line-filled and gray-filled areas indicate
 approximate ranges of electron temperature for the FIZ and the PIZ,
 respectively.}  \label{fig:collision_excite}
\end{figure}

The occurrence of the \ion{Mg}{2} emissivity peak in the FIZ is
explained as follows. First of all, since the Mg abundance is several
orders smaller than H and He, the recombination of electrons with
Mg$^{2+}$ into the excited state of Mg$^+$ is superseded by collisional
excitation of Mg$^+$ with electrons. Consider that an ion in the lower
level 1 is excited to the upper level 2 by collision with electrons.
Then, the excitation rate is written as $n_\mathrm{e}n_1 q_{12}$, where
$n_\mathrm{e}$ is the electron density, $n_1$ is the ion density in
level 1, and $q_{12}$ is the rate coefficient for collisional excitation
from level 1 to level 2 given by
\begin{equation}
 q_{12} = \frac{8.629\times10^{-6}}{T_\mathrm{e}^{1/2}}
  \frac{\Upsilon(1,2)}{\omega_1} \exp(-\chi/kT_\mathrm{e}),
\end{equation}
where $T_\mathrm{e}$ is the electron temperature, $\omega_1$ is a
statistical weight of level 1, $\chi$ is the excitation energy, and
$\Upsilon(1,2)$ is a collision strength that varies slowly with
$T_\mathrm{e}$ (\citealt{2006agna.book.....O}). Figure
\ref{fig:collision_excite} shows $q_{12}$ as a function of
$T_\mathrm{e}$ for the \ion{Mg}{2} $\lambda2798$ emission line.  It is
evident that the rate coefficient at $T_\mathrm{e}\sim 10^4$ K in the
FIZ is about two orders larger than at $T_\mathrm{e}\sim 5\times10^{3}$
K in the PIZ. In addition, the self absorption of line photons by
abundant Mg$^+$ from the PIZ diminishes the emergent \ion{Mg}{2} flux in
that zone. As a result, the \ion{Mg}{2} line emissivity peaks in the
FIZ.

\subsection{Dependence on non-abundance parameters}
\begin{figure*}[t]
  \epsscale{1.0}
 \plotone{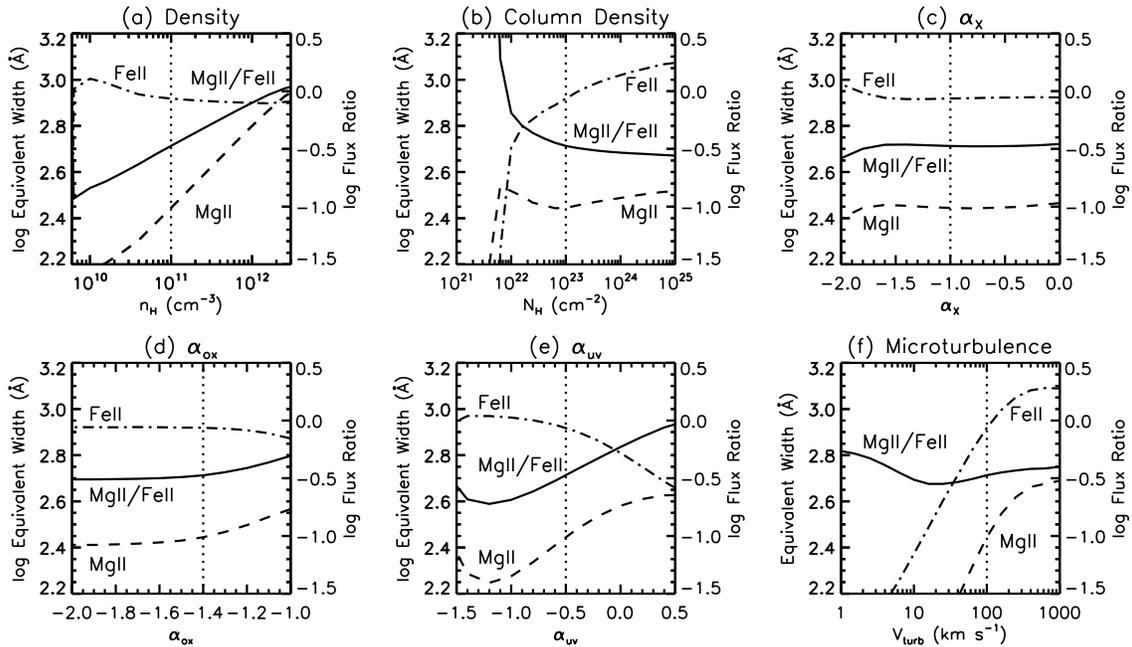}
 \caption{Dependence of the emission line on non-abundance parameters of
 the BLR cloud. Each panel shows the dependence on each parameter: ($a$)
 hydrogen gas density $n_\mathrm{H}$, ($b$) hydrogen column density
 $N_\mathrm{H}$, ($c$) X-ray power-law index $\alpha_{\mathrm{x}}$,
 ($d$) optical to X-ray spectral index $\alpha_{\mathrm{ox}}$, ($e$) UV
 power-law index $\alpha_{\mathrm{uv}}$ and ($f$) microturbulence
 $v_{\mathrm{turb}}$.  The dash-dotted, dashed, and thick lines indicate
 EW(\ion{Fe}{2}), EW(\ion{Mg}{2}), and \ion{Mg}{2}/\ion{Fe}{2},
 respectively. The vertical dotted line indicates the fiducial value of
 each parameter.}  \label{fig:ew_dependence}
\end{figure*}

The calculated results for parameter dependence of emission lines are
summarized in Figure \ref{fig:ew_dependence}. Each panel of this figure
shows how the EWs of \ion{Mg}{2} and \ion{Fe}{2}, as well as the
\ion{Mg}{2}/\ion{Fe}{2} flux ratio, change with the change of one
particular parameter, while other parameters are fixed to their fiducial
values of the baseline model.  In the following, we explain the results
for changing the hydrogen gas density $n_\mathrm{H}$, the column density
$N_\mathrm{H}$, the SED of the incident continuum, and the
microturbulence $v_{\mathrm{turb}}$.

\begin{figure}[t]
  \epsscale{1.0} \plotone{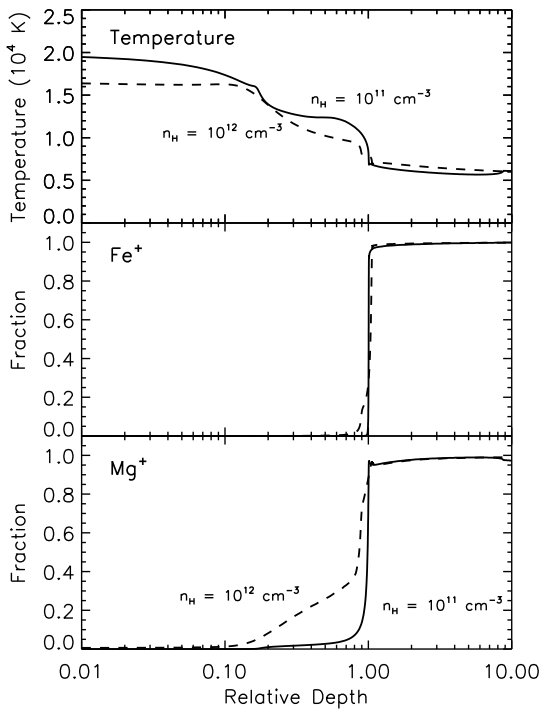}
 \caption{Change of electron temperature and ionization structure for
 different values of the BLR gas density. The horizontal axis indicates
 the relative depth of a cloud that is normalized at the ionization
 front.  The vertical axes indicate electron temperature ({\it top
 panel}) and fractions of Fe$^+$ ({\it middle panel}) and Mg$^+$ ({\it
 bottom panel}). The solid and dashed lines shown on each panel are for
 $n_\mathrm{H}=10^{11}\ \mathrm{cm^{-3}}$ and $10^{12}\
 \mathrm{cm^{-3}}$, respectively.}  \label{fig:ionfract_compare}
\end{figure}
The $n_\mathrm{H}$ dependence is shown in panel $a$ of Figure
\ref{fig:ew_dependence}.  Note that $n_\mathrm{H}$ and $U$ are varied
simultaneously to keep their product constant ($n_\mathrm{H}U=10^{9.5}\
\mathrm{cm^{-3}}$), according to the BLR reverberation constraint (see
Section 4.1).  As can clearly be seen, EW(\ion{Mg}{2}) increases with
increasing $n_\mathrm{H}$, while EW(\ion{Fe}{2}) hardly changes with
it. The physical reason for this is clear.  In the FIZ, as
$n_\mathrm{H}$ increases, $T_\mathrm{e}$ decreases and the fraction of
Mg$^+$ ions increases (see Figure \ref{fig:ionfract_compare}).  This is
due to enhanced recombination of $\mathrm{Mg}^{2+}+\mathrm{e}^{-}
\rightarrow \mathrm{Mg}^{+}$ for lower $T_\mathrm{e}$ because its rate
is proportional to ${T_\mathrm{e}}^{-1/2}$. On the other hand, in the
PIZ, $T_\mathrm{e}$ hardly changes, so that the fraction of Fe$^{+}$ and
Mg$^{+}$ ions keeps almost unchanged.  As a result, the strength of the
\ion{Mg}{2} emission line from the FIZ increases, while that of the
\ion{Fe}{2} emission line from the PIZ is kept almost constant.

The $N_\mathrm{H}$ dependence is shown in panel $b$ of Figure
\ref{fig:ew_dependence}.  For $N_\mathrm{H} \lesssim 10^{22}\
\mathrm{cm^{-2}}$, both EW(\ion{Mg}{2}) and EW(\ion{Fe}{2}) change
drastically. The \ion{Fe}{2} emission line does not emerge from the PIZ,
because the PIZ is not formed for such lower $N_\mathrm{H}$ and the gas
is fully ionized throughout the cloud.  Furthermore, below
$N_\mathrm{H}\sim 10^{22}\ \mathrm{cm^{-2}}$, the \ion{Mg}{2} emission
line does not emerge from the FIZ, because the FIZ is not extended
deeply enough to reach the \ion{Mg}{2} line-forming region. The fact
that most quasars show both \ion{Mg}{2} and \ion{Fe}{2} emission lines
suggests that BLR clouds generally have $N_\mathrm{H} \gtrsim 10^{22}\
\mathrm{cm^{-2}}$. As $N_\mathrm{H}$ increases beyond $10^{22}\
\mathrm{cm^{-2}}$, the size of the PIZ increases. As a result, the
strength of the \ion{Fe}{2} emission line from the PIZ increases, while
the strength of the \ion{Mg}{2} emission line from the FIZ is
unaffected.  Note that \cite{2009ApJ...707L..82F} similarly argued for
the $N_\mathrm{H}$ dependence of the emission line for \ion{Fe}{2}, but
not \ion{Mg}{2}.

The results for varying the X-ray power-law index $\alpha_{\mathrm{x}}$
and the spectral index $\alpha_{\mathrm{ox}}$, both of which effectively
determine the strength of the SED at $\gtrsim 1$ keV, are shown in
panels $c$ and $d$ of Figure \ref{fig:ew_dependence}.  Our
photoionization calculations show that both EW(\ion{Mg}{2}) and
EW(\ion{Fe}{2}) do not change very much with varying
$\alpha_{\mathrm{x}}$ and $\alpha_{\mathrm{ox}}$.  In particular, the
changes of EW(\ion{Mg}{2}) and \ion{Mg}{2}/\ion{Fe}{2} with the X-ray
SED are too small to reproduce their observed ranges.  Similar results
are also reported by \cite{2010ApJ...721.1835S}.  Because incident X-ray
photons are widely assumed as the main heating source in the PIZ, it may
seem strange that the \ion{Fe}{2} emission depends little on the X-ray
SED.  In fact, our calculations indicate that harder SED gives rise to
larger electron density and hence more cooling by Fe$^{+}$ ions
especially at large cloud depth in the PIZ. However, the emissivities of
those additionally created UV \ion{Fe}{2} photons at large cloud depth
are quite small due to their large optical depths there (see Figure
\ref{fig:emissivity}).  As a result, the emissivity of UV \ion{Fe}{2}
emission line as a whole is little affected by the X-ray SED of an
illuminating source.

The dependence on the UV power-law index $\alpha_{\mathrm{uv}}$ is shown
in panel $e$ of Figure \ref{fig:ew_dependence}.  As indicated in the
figure, EW(\ion{Fe}{2}) decreases with $\alpha_{\mathrm{uv}}$.  This is
mainly due to less incident photons having the energy less than 13.6 eV
for larger $\alpha_{\mathrm{uv}}$.  Those low-energy photons can
penetrate the FIZ and photoexcite Fe$^{+}$ ions in the PIZ.  Therefore,
continuum pumping of Fe$^{+}$ ions and the resulting \ion{Fe}{2}
emission lines decrease when $\alpha_{\mathrm{uv}}$ increases.  On the
other hand, our photoionization calculations indicate that
EW(\ion{Mg}{2}) increases with $\alpha_{\mathrm{uv}}$.  Moreover, we
find that the line flux of \ion{Mg}{2} hardly changes with
$\alpha_{\mathrm{uv}}$; this result is understandable, because
\ion{Mg}{2} emission lines are mainly created at the FIZ which is
transparent to those low energy photons with $E \lesssim 13.6$ eV.
Thus, the increase of EW(\ion{Mg}{2}) with $\alpha_{\mathrm{uv}}$ is
mainly due to the decrease of continuum flux at 3000\AA.

The $v_\mathrm{turb}$ dependence is shown in panel $f$ of Figure
\ref{fig:ew_dependence}. Larger $v_\mathrm{turb}$ gives rise to larger
EW(\ion{Mg}{2}) and EW(\ion{Fe}{2}); this is because the local velocity
difference in the moving medium decreases the self-absorption of the
line photons, which causes them to escape more easily from the
cloud. However, by calculating the \ion{Mg}{2}/\ion{Fe}{2} flux ratio,
such $v_\mathrm{turb}$ dependence for each of the EWs is largely
cancelled out.  The resultant ratio hardly changes with
$v_{\mathrm{turb}}$.

\section{Discussion}

\subsection{Density dependence of the \ion{Mg}{2}/\ion{Fe}{2} flux ratio}

The assumption that the \ion{Mg}{2}/\ion{Fe}{2} flux ratio is a
first-order proxy for the Mg/Fe abundance ratio would hold only if the
\ion{Mg}{2} and \ion{Fe}{2} emission lines were found to originate from
the same region owing to their similar ionization energies (e.g.,
\citealt{1999ARA&A..37..487H}).  However, the result of our
photoionization calculations shown in Figure \ref{fig:emissivity} is
highly suggestive of a difference in the emergent zone between the
\ion{Mg}{2} and \ion{Fe}{2} emission lines; \ion{Mg}{2} flux comes
mostly from the FIZ, while \ion{Fe}{2} flux comes from the PIZ. As can
be seen from Figure \ref{fig:ionfract_compare}, the ionization structure
of a BLR cloud depends on the non-abundance parameter of $n_\mathrm{H}$,
and more importantly, this $n_\mathrm{H}$ dependence for the FIZ is
different from that for the PIZ. Thus, our result invalidates the
assumption of \ion{Mg}{2}/\ion{Fe}{2} $\propto$ Mg/Fe, and indicates
that the \ion{Mg}{2}/\ion{Fe}{2} flux ratio strongly depends on some
non-abundance parameter other than the Mg/Fe abundance ratio.  Our
result therefore supports similar arguments previously given by other
authors (e.g., \citealt{2011ApJ...736...86D};
\citealt{2011ApJ...739...56D}).

The Mg/Fe abundance ratio can then be derived from the
\ion{Mg}{2}/\ion{Fe}{2} flux ratio {\it after} the effects of some
non-abundance parameter, yet to be identified as a main driver of the
existing correlations among observable quantities of quasars, have been
properly corrected. Comparison of Figures \ref{fig:fluxratio_ew_edd} and
\ref{fig:ew_dependence} hints at a way of identifying such a parameter.
Figure \ref{fig:fluxratio_ew_edd} indicates that \ion{Mg}{2}
emission-line flux correlates with $L_{\mathrm{bol}}/L_{\mathrm{Edd}}$,
but \ion{Fe}{2} emission-line flux does not.  From the result of our
photoionization calculations in Figure \ref{fig:ew_dependence}, those
trends with $L_{\mathrm{bol}}/L_{\mathrm{Edd}}$ are fully reproduced
solely by changing $n_\mathrm{H}$, when an anticorrelation exists
between $n_\mathrm{H}$ and $L_{\mathrm{bol}}/L_{\mathrm{Edd}}$.  On the
other hand, those trends are not reproduced by changing either
$N_\mathrm{H}$, the SED, or $v_\mathrm{turb}$. Therefore, it is
reasonable to proceed with a hypothesis that the non-abundance parameter
to be identified is mainly the BLR gas density that anticorrelates with
the Eddington ratio, and the observed diversity of \ion{Mg}{2} flux is
partly attributed to object to object variation of the BLR gas density
in quasars.

\begin{figure}[t]
 \epsscale{1.0} \plotone{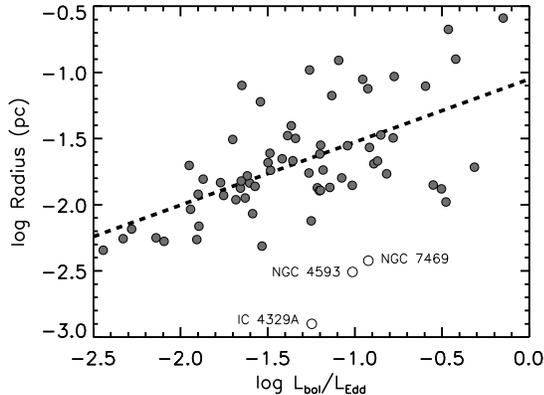}
 \caption{Distance to BLR clouds from the central source plotted against
 the Eddington ratio. The filled and open circles show the data taken
 from \cite{2009ApJ...697..160B}, and their black hole masses are
 retrieved from \cite{2004ApJ...613..682P}. The dashed line indicates
 the linear regression line.  The open circles are removed from the
 regression analysis because of poor-quality lag measurement (IC 4329A
 and NGC 4593) and contamination from the nuclear starburst ring (NGC
 7469).}  \label{fig:radius_edd}
\end{figure}
Our hypothesis of anticorrelation between $n_\mathrm{H}$ and
$L_{\mathrm{bol}}/L_{\mathrm{Edd}}$ could be supported by the following
discussion. For our sample of SDSS quasars, we obtain EW(\ion{Mg}{2})
$\propto (L_{\mathrm{bol}}/L_{\mathrm{Edd}})^{-0.30}$ (Figure
\ref{fig:fluxratio_ew_edd} and Table \ref{tab:correlation}), and from
our photoionization calculations, we find EW(\ion{Mg}{2}) $\propto
{n_\mathrm{H}}^{0.34}$ (Figure \ref{fig:ew_dependence}). These two
relations are combined to give $n_\mathrm{H} \propto
(L_{\mathrm{bol}}/L_{\mathrm{Edd}})^{-0.88}$.  On the other hand,
\cite{1989ApJ...347..640R} argued that the BLR gas density decreases
with the increasing radial distance of the cloud away from the AGN
center, i.e., $n_\mathrm{H} \propto R^{-s}\ (1 < s < 2.5)$.  Therefore,
our hypothesis requires $R \propto
(L_{\mathrm{bol}}/L_{\mathrm{Edd}})^b$, provided $b=0.88/s$.  We can
check this requisite using the reverberation mapping data given in
\cite{2009ApJ...697..160B}.\footnote{The latest work on the
radius--luminosity relationship by reverberation mapping measurement is
given by \cite{2013ApJ...767..149B}, where six AGNs are newly added to
those in \cite{2009ApJ...697..160B}. However, their black hole masses
are not given, and it is not possible to estimate their Eddington
ratios.  Therefore, in this study, we use the data of
\cite{2009ApJ...697..160B} for which all black hole masses are available
from \cite{2004ApJ...613..682P}.} The result is shown in Figure
\ref{fig:radius_edd}, where a positive correlation is clearly seen.  A
linear regression analysis gives $b=0.48$, leading to $s = 0.88/0.48
\sim 1.8$ within the limits of $s$ in accordance with
\cite{1989ApJ...347..640R}.  In this way, our hypothesis of
anticorrelation between $n_\mathrm{H}$ and
$L_{\mathrm{bol}}/L_{\mathrm{Edd}}$ is well confirmed by the data.

\subsection{Heavy-element abundance diagnostics with \ion{Mg}{2} and \ion{Fe}{2}
  emission lines}

\begin{figure}[t]
  \epsscale{1.0} \plotone{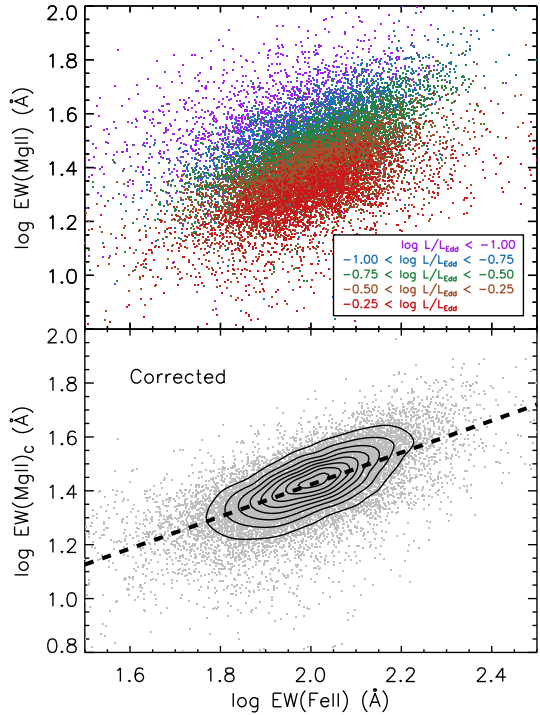}
  \caption{EW(\ion{Mg}{2}) versus EW(\ion{Fe}{2}) diagram. {\it Top:}
  The measured EW(\ion{Mg}{2})s for our sample of SDSS quasars are
  plotted against the measured EW(\ion{Fe}{2})s. Quasars are colored
  differently according to the different ranges of the Eddington ratio.
  {\it Bottom:} The EW(\ion{Mg}{2})s corrected for the density effect
  using equation (\ref{eq:density_correct}) are plotted against the
  measured EW(\ion{Fe}{2})s as gray dots. The contours indicate local
  data point densities with a grid size of $\Delta=0.01$ dex. The dashed
  line indicates a linear regression line.}
  \label{fig:ew_mgii_feii_eddcorrect}
\end{figure}
Establishing a method of deriving heavy-element abundances of a BLR
cloud is the most important problem in chemical evolution studies using
quasars.  For a long time, the \ion{Fe}{2}/\ion{Mg}{2} flux ratio has
been used as a proxy for the Fe/Mg abundance ratio.  However, as
described in the previous section, we have reached the conclusion that
the strength of the \ion{Mg}{2} emission line strongly depends on the
non-abundance parameter of the BLR gas density.  Accordingly, the
\ion{Mg}{2}/\ion{Fe}{2} flux ratio, unless corrected for the density
effect, can never be used as a measure of the Mg/Fe abundance ratio.  In
this section, we describe a conversion of flux to abundance for quasars.

The first step is to correct EW(MgII) for the density effect, by scaling
the observed EW(MgII) values of all quasars to those for the reference
value of the Eddington ratio. The corrected EW(MgII) is given by
\begin{equation}
 \mathrm{EW(MgII)}_{\mathrm{c}} = \mathrm{EW(MgII)} \left( \frac{\langle
						     L_{\mathrm{bol}}/L_{\mathrm{Edd}}\rangle}{L_{\mathrm{bol}}/L_{\mathrm{Edd}}}
						    \right)^{-0.30},
 \label{eq:density_correct}
\end{equation}
where $\langle L_{\mathrm{bol}}/L_{\mathrm{Edd}}\rangle =10^{-0.55}$ is
adopted as a median value of the measured Eddington ratios of the sample
quasars (see Figure \ref{fig:measure}).  Figure
\ref{fig:ew_mgii_feii_eddcorrect} shows the observed EW(\ion{Mg}{2})
versus EW(\ion{Fe}{2}) diagram in the upper panel, and the corrected
EW(\ion{Mg}{2})$_{\mathrm{c}}$ versus EW(\ion{Fe}{2}) diagram in the
lower panel.  Obviously, the scatter in the corrected EW(\ion{Mg}{2})
data is, as a matter of course, decreased.

\begin{figure}[t]
  \epsscale{1.0} \plotone{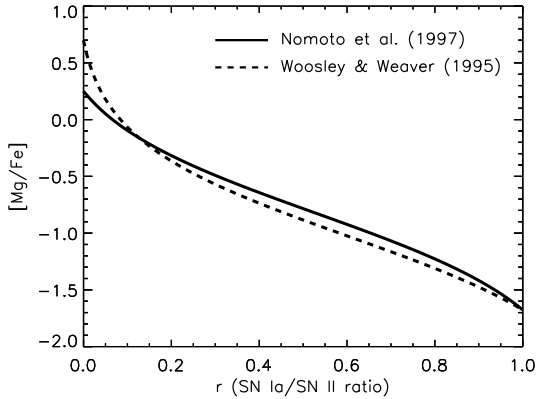}
 \caption{Logarithmic abundance ratio [Mg/Fe] as a function of $r$ which
 represents the fraction of the SN Ia contribution in the abundance
 pattern. The solid and dashed lines indicate the results where the
 abundance pattern of SN II ejecta has been taken from
 \cite{1997NuPhA.616...79N} and \cite{1995ApJS..101..181W},
 respectively.}  \label{fig:mgfe_sn_ratio}
\end{figure}

\begin{turnpage}
 \begin{deluxetable*}{lrrrrrrrrrrrr}
  \tabletypesize{\scriptsize}
  \tablecaption{Elemental abundance pattern for different mixtures of SN Ia
  and SN II: \cite{1997NuPhA.616...79N} \label{tab:abuptn_nomoto}}
  \tablewidth{0pt}
  \tablehead{
  \colhead{Mg/Fe\tablenotemark{$\dagger$}} & \colhead{1/10} &
  \colhead{1/8} & \colhead{1/6} & \colhead{1/4} & \colhead{1/2} &
  \colhead{1} & \colhead{1.781} & \colhead{2} & \colhead{4} &
  \colhead{6} & \colhead{8} & \colhead{10} \\
  $(r, M_{\mathrm{min}}/M_{\odot})$ & \colhead{$(0.653, 10)$}
  & \colhead{$(0.585, 10)$} & \colhead{$(0.495, 10)$} &
  \colhead{$(0.372, 10)$} & \colhead{$(0.191, 10)$} & \colhead{$(0.066,
  10)$} & \colhead{$(0, 10)$} & \colhead{$(0, 10.53)$} &
  \colhead{$(0, 14.07)$} & \colhead{$(0, 16.45)$} &
  \colhead{$(0, 19.20)$} &
  \colhead{$(0, 24.78)$}
  }
  \startdata
  C	& $-4.91$	& $-4.88$	& $-4.84$	& $-4.76$	& $-4.58$	& $-4.36$	& $-4.16$	& $-4.05$	& $-3.68$	& $-3.56$	& $-3.53$	& $-3.50$	\\
  N	& $-8.52$	& $-8.41$	& $-8.27$	& $-8.08$	& $-7.76$	& $-7.45$	& $-7.20$	& $-6.14$	& $-5.47$	& $-5.52$	& $-5.65$	& $-5.89$	\\
  O	& $-4.03$	& $-3.94$	& $-3.81$	& $-3.64$	& $-3.34$	& $-3.04$	& $-2.79$	& $-2.74$	& $-2.45$	& $-2.27$	& $-2.14$	& $-2.00$	\\
  F	& $-12.92$	& $-12.87$	& $-12.79$	& $-12.67$	& $-12.43$	& $-12.16$	& $-11.93$	& $-11.93$	& $-11.77$	& $-11.65$	& $-12.05$	& $-12.02$	\\
  Ne	& $-5.09$	& $-4.98$	& $-4.84$	& $-4.65$	& $-4.33$	& $-4.03$	& $-3.77$	& $-3.73$	& $-3.44$	& $-3.26$	& $-3.15$	& $-3.06$	\\
  Na	& $-6.70$	& $-6.58$	& $-6.44$	& $-6.25$	& $-5.93$	& $-5.62$	& $-5.36$	& $-5.33$	& $-5.05$	& $-4.87$	& $-4.76$	& $-4.62$	\\
  Mg	& $-5.40$	& $-5.30$	& $-5.18$	& $-5.00$	& $-4.70$	& $-4.40$	& $-4.15$	& $-4.10$	& $-3.80$	& $-3.62$	& $-3.50$	& $-3.40$	\\
  Al	& $-6.37$	& $-6.27$	& $-6.15$	& $-5.97$	& $-5.67$	& $-5.37$	& $-5.12$	& $-5.06$	& $-4.76$	& $-4.58$	& $-4.45$	& $-4.32$	\\
  Si	& $-4.81$	& $-4.79$	& $-4.76$	& $-4.69$	& $-4.55$	& $-4.35$	& $-4.16$	& $-4.13$	& $-3.91$	& $-3.76$	& $-3.66$	& $-3.52$	\\
  P	& $-7.26$	& $-7.20$	& $-7.12$	& $-7.00$	& $-6.76$	& $-6.49$	& $-6.26$	& $-6.21$	& $-5.92$	& $-5.76$	& $-5.64$	& $-5.48$	\\
  S	& $-5.16$	& $-5.15$	& $-5.13$	& $-5.10$	& $-5.02$	& $-4.89$	& $-4.74$	& $-4.69$	& $-4.43$	& $-4.29$	& $-4.19$	& $-4.04$	\\
  Cl	& $-7.87$	& $-7.85$	& $-7.83$	& $-7.77$	& $-7.65$	& $-7.47$	& $-7.29$	& $-7.25$	& $-7.01$	& $-6.85$	& $-6.74$	& $-6.57$	\\
  Ar	& $-5.94$	& $-5.93$	& $-5.92$	& $-5.88$	& $-5.80$	& $-5.67$	& $-5.52$	& $-5.46$	& $-5.19$	& $-5.05$	& $-4.94$	& $-4.78$	\\
  K	& $-8.19$	& $-8.17$	& $-8.14$	& $-8.09$	& $-7.96$	& $-7.78$	& $-7.60$	& $-7.55$	& $-7.29$	& $-7.13$	& $-7.01$	& $-6.83$	\\
  Ca	& $-6.10$	& $-6.09$	& $-6.08$	& $-6.04$	& $-5.96$	& $-5.84$	& $-5.69$	& $-5.64$	& $-5.38$	& $-5.25$	& $-5.15$	& $-5.00$	\\
  Sc	& $-10.80$	& $-10.78$	& $-10.74$	& $-10.68$	& $-10.53$	& $-10.33$	& $-10.13$	& $-10.08$	& $-9.80$	& $-9.63$	& $-9.51$	& $-9.32$	\\
  Ti	& $-7.85$	& $-7.83$	& $-7.81$	& $-7.76$	& $-7.64$	& $-7.47$	& $-7.29$	& $-7.29$	& $-7.21$	& $-7.14$	& $-7.09$	& $-7.09$	\\
  V	& $-8.61$	& $-8.61$	& $-8.61$	& $-8.60$	& $-8.59$	& $-8.55$	& $-8.51$	& $-8.48$	& $-8.30$	& $-8.21$	& $-8.16$	& $-8.05$	\\
  Cr	& $-6.53$	& $-6.52$	& $-6.52$	& $-6.51$	& $-6.49$	& $-6.44$	& $-6.38$	& $-6.36$	& $-6.20$	& $-6.10$	& $-6.05$	& $-5.96$	\\
  Mn	& $-6.46$	& $-6.47$	& $-6.47$	& $-6.49$	& $-6.54$	& $-6.66$	& $-6.95$	& $-6.92$	& $-6.75$	& $-6.66$	& $-6.63$	& $-6.53$	\\
  Fe	& $-4.50$	& $-4.50$	& $-4.50$	& $-4.50$	& $-4.50$	& $-4.50$	& $-4.50$	& $-4.50$	& $-4.50$	& $-4.50$	& $-4.50$	& $-4.50$	\\
  Co	& $-7.39$	& $-7.40$	& $-7.40$	& $-7.41$	& $-7.43$	& $-7.49$	& $-7.59$	& $-7.60$	& $-7.65$	& $-7.66$	& $-7.66$	& $-7.94$	\\
  Ni	& $-5.26$	& $-5.26$	& $-5.27$	& $-5.28$	& $-5.33$	& $-5.44$	& $-5.72$	& $-5.72$	& $-5.72$	& $-5.76$	& $-5.78$	& $-5.86$	\\
  Cu	& $-9.97$	& $-9.96$	& $-9.93$	& $-9.88$	& $-9.75$	& $-9.58$	& $-9.40$	& $-9.40$	& $-9.41$	& $-9.36$	& $-9.31$	& $-9.68$	\\
  Zn	& $-8.84$	& $-8.82$	& $-8.79$	& $-8.73$	& $-8.59$	& $-8.41$	& $-8.21$	& $-8.23$	& $-8.29$	& $-8.32$	& $-8.30$	& $-8.73$	
  \enddata
  
  \tablecomments{The values are the logarithmic number density of each
  element relative to hydrogen. All heavy elements are scaled so that
  the resultant [Fe/H] equals 0.}

  \tablenotetext{$\dagger$}{$\mathrm{Mg/Fe} \equiv 10^{\mathrm{[Mg/Fe]}}$.}
 \end{deluxetable*}
\end{turnpage}
 
\begin{turnpage}
 \begin{deluxetable*}{lrrrrrrrrrrrr}
  \tabletypesize{\scriptsize}
  \tablecaption{Elemental abundance pattern for different mixtures of SN Ia
  and SN II: \cite{1995ApJS..101..181W} \label{tab:abuptn_woosley}}
  \tablewidth{0pt}
  \tablehead{
  \colhead{Mg/Fe\tablenotemark{$\dagger$}} & \colhead{1/10} & \colhead{1/8} & \colhead{1/6} &
  \colhead{1/4} & \colhead{1/2} & \colhead{1} & \colhead{2} & \colhead{4} & \colhead{5.280} &
  \colhead{6} & \colhead{8} & \colhead{10} \\
  $(r, M_{\mathrm{min}}/M_{\odot})$ & \colhead{$(0.582, 10)$} & \colhead{$(0.513, 10)$}
  & \colhead{$(0.427, 10)$} & \colhead{$(0.318, 10)$} &
  \colhead{$(0.174, 10)$} & \colhead{$(0.084, 10)$} & \colhead{$(0.032,
  10)$} & \colhead{$(0.005, 10)$} & \colhead{$(0, 10)$} & \colhead{$(0,
  13.73)$} & \colhead{$(0, 20.57)$} & \colhead{$(0, 24.49)$}
  }
  \startdata
  C	& $-4.57$	& $-4.49$	& $-4.38$	& $-4.22$	& $-3.94$	& $-3.65$	& $-3.34$	& $-3.04$	& $-2.96$	& $-2.91$ & $-2.89$ & $-2.85$ \\
  N	& $-5.27$	& $-5.15$	& $-5.01$	& $-4.81$	& $-4.49$	& $-4.18$	& $-3.87$	& $-3.56$	& $-3.47$	& $-3.47$ & $-3.41$ & $-3.34$ \\
  O	& $-3.96$	& $-3.86$	& $-3.73$	& $-3.55$	& $-3.25$	& $-2.95$	& $-2.64$	& $-2.33$	& $-2.25$	& $-2.15$ & $-2.03$ & $-1.99$ \\
  F	& $-8.46$	& $-8.34$	& $-8.20$	& $-8.00$	& $-7.68$	& $-7.37$	& $-7.06$	& $-6.75$	& $-6.66$	& $-6.60$ & $-6.49$ & $-6.39$ \\
  Ne	& $-4.95$	& $-4.84$	& $-4.70$	& $-4.51$	& $-4.19$	& $-3.88$	& $-3.57$	& $-3.26$	& $-3.18$	& $-3.10$ & $-3.00$ & $-2.89$ \\
  Na	& $-6.60$	& $-6.48$	& $-6.34$	& $-6.14$	& $-5.82$	& $-5.52$	& $-5.20$	& $-4.90$	& $-4.81$	& $-4.75$ & $-4.67$ & $-4.54$ \\
  Mg	& $-5.40$	& $-5.30$	& $-5.18$	& $-5.00$	& $-4.70$	& $-4.40$	& $-4.09$	& $-3.79$	& $-3.70$	& $-3.62$ & $-3.50$ & $-3.40$ \\
  Al	& $-6.32$	& $-6.22$	& $-6.10$	& $-5.92$	& $-5.61$	& $-5.31$	& $-5.00$	& $-4.70$	& $-4.61$	& $-4.52$ & $-4.36$ & $-4.26$ \\
  Si	& $-4.75$	& $-4.71$	& $-4.65$	& $-4.56$	& $-4.36$	& $-4.13$	& $-3.85$	& $-3.57$	& $-3.48$	& $-3.43$ & $-3.43$ & $-3.60$ \\
  P	& $-7.07$	& $-6.99$	& $-6.88$	& $-6.72$	& $-6.44$	& $-6.15$	& $-5.85$	& $-5.55$	& $-5.46$	& $-5.38$ & $-5.28$ & $-5.33$ \\
  S	& $-5.07$	& $-5.04$	& $-4.99$	& $-4.91$	& $-4.73$	& $-4.50$	& $-4.23$	& $-3.95$	& $-3.87$	& $-3.83$ & $-3.89$ & $-4.21$ \\
  Cl	& $-7.51$	& $-7.43$	& $-7.32$	& $-7.17$	& $-6.89$	& $-6.60$	& $-6.29$	& $-5.99$	& $-5.91$	& $-5.86$ & $-5.83$ & $-6.14$ \\
  Ar	& $-5.85$	& $-5.82$	& $-5.77$	& $-5.68$	& $-5.49$	& $-5.26$	& $-4.99$	& $-4.71$	& $-4.62$	& $-4.60$ & $-4.66$ & $-5.00$ \\
  K	& $-7.79$	& $-7.71$	& $-7.60$	& $-7.44$	& $-7.15$	& $-6.86$	& $-6.56$	& $-6.26$	& $-6.17$	& $-6.19$ & $-6.19$ & $-6.55$ \\
  Ca	& $-6.03$	& $-6.01$	& $-5.97$	& $-5.89$	& $-5.72$	& $-5.51$	& $-5.25$	& $-4.98$	& $-4.89$	& $-4.88$ & $-4.98$ & $-5.26$ \\
  Sc	& $-10.16$	& $-10.06$	& $-9.93$	& $-9.75$	& $-9.45$	& $-9.15$	& $-8.84$	& $-8.54$	& $-8.45$	& $-8.41$ & $-8.33$ & $-8.61$ \\
  Ti	& $-7.85$	& $-7.83$	& $-7.81$	& $-7.76$	& $-7.64$	& $-7.47$	& $-7.25$	& $-7.00$	& $-6.92$	& $-6.92$ & $-6.91$ & $-6.95$ \\
  V	& $-8.60$	& $-8.59$	& $-8.58$	& $-8.56$	& $-8.51$	& $-8.42$	& $-8.28$	& $-8.09$	& $-8.02$	& $-8.02$ & $-7.98$ & $-7.96$ \\
  Cr	& $-6.53$	& $-6.53$	& $-6.53$	& $-6.52$	& $-6.51$	& $-6.48$	& $-6.42$	& $-6.33$	& $-6.30$	& $-6.30$ & $-6.30$ & $-6.30$ \\
  Mn	& $-6.45$	& $-6.45$	& $-6.45$	& $-6.45$	& $-6.46$	& $-6.47$	& $-6.50$	& $-6.55$	& $-6.58$	& $-6.59$ & $-6.56$ & $-6.51$ \\
  Fe	& $-4.50$	& $-4.50$	& $-4.50$	& $-4.50$	& $-4.50$	& $-4.50$	& $-4.50$	& $-4.50$	& $-4.50$	& $-4.50$ & $-4.50$ & $-4.50$ \\
  Co	& $-7.36$	& $-7.35$	& $-7.34$	& $-7.32$	& $-7.26$	& $-7.15$	& $-6.99$	& $-6.79$	& $-6.72$	& $-6.66$ & $-6.58$ & $-6.53$ \\
  Ni	& $-5.23$	& $-5.23$	& $-5.23$	& $-5.22$	& $-5.20$	& $-5.15$	& $-5.08$	& $-4.95$	& $-4.91$	& $-5.18$ & $-5.23$ & $-5.36$ \\
  Cu	& $-8.70$	& $-8.59$	& $-8.45$	& $-8.26$	& $-7.94$	& $-7.63$	& $-7.32$	& $-7.01$	& $-6.93$	& $-6.83$ & $-6.69$ & $-6.65$ \\
  Zn	& $-8.07$	& $-7.96$	& $-7.83$	& $-7.65$	& $-7.34$	& $-7.03$	& $-6.72$	& $-6.42$	& $-6.33$	& $-6.23$ & $-6.06$ & $-6.00$
  \enddata
  
  \tablecomments{The values are the logarithmic number density of each
  element relative to hydrogen. All heavy elements are scaled so that
  the resultant [Fe/H] equals 0.}

  \tablenotetext{$\dagger$}{$\mathrm{Mg/Fe} \equiv 10^{\mathrm{[Mg/Fe]}}$.}
 \end{deluxetable*}
\end{turnpage}

The second step is to bring about one-to-one correspondence between the
(Mg/Fe, Fe/H) abundance pair and the (EW(\ion{Mg}{2})$_{\mathrm{c}}$,
EW(\ion{Fe}{2})) pair, by running the {\sc Cloudy} simulation code with
other non-abundance parameters fixed to the fiducial values given in
Table \ref{tab:baseline}.  For realistic simulations with the
heavy-element abundance pattern of stellar origin, we mix the yield
patterns of SN Ia and SN II, following \cite{1995MNRAS.277..945T}. The
ejecta masses of heavy elements for SN Ia are adopted from the W7 model
in \cite{1997NuPhA.621..467N}, and those for SN II are adopted from
\cite{1997NuPhA.616...79N} and \cite{1995ApJS..101..181W}.  Since the
ejecta mass of each element for SN II depends on the mass of the
progenitor star, we calculate the mean ejecta mass weighted by an
initial mass function (IMF) of stars. Then the mean ejecta mass for the
$i$th heavy element is given by
\begin{equation}
 M_{i,\mathrm{II}} = \frac{\displaystyle \int_{m_\mathrm{min}}^{m_\mathrm{max}}
  M_{i,\mathrm{II}}(m)m^{-(1+\mu)}dm}{\displaystyle \int_{m_\mathrm{min}}^{m_\mathrm{max}}
  m^{-(1+\mu)}dm},  \label{eq:mean_ejecta_mass}
\end{equation}
where $m$ is the mass of the progenitor star, $M_i(m)$ is the $i$th
heavy-element mass produced in a main-sequence star of mass $m$,
$m_\mathrm{min}$ and $m_\mathrm{max}$ are the lower and upper progenitor
masses, and $\mu$ is the power index of the IMF. In this paper, we adopt
$(m_\mathrm{min}, m_\mathrm{max})=(10M_\odot, 50M_\odot)$ and
$\mu=1.35$, which represents the Salpeter IMF.
 
We introduce a parameter $r$ for the mass fraction contributed by SN Ia
per unit mass of all heavy elements. Then, the heavy-element abundance
pattern is written as
\begin{equation}
 x_i = r x_{i,\mathrm{Ia}} + (1-r) x_{i, \mathrm{II}} ,
\end{equation}
where $x_{i,\mathrm{Ia}} \equiv M_{i,\mathrm{Ia}}/\sum_{i=6}^{30}
M_{i,\mathrm{Ia}}$ and $x_{i,\mathrm{II}} \equiv
M_{i,\mathrm{II}}/\sum_{i=6}^{30} M_{i,\mathrm{II}}$. Note that since
the elements given in \cite{1997NuPhA.616...79N} are from C to Zn, the
index $i$ in the above summation starts from 6 and ends at 30. The
abundance ratio between Mg and Fe is equal to
$x_\mathrm{Mg}/x_\mathrm{Fe}$, and in the following, for the sake of
brevity, we denote Mg/Fe as the Mg/Fe abundance ratio in units of the
solar ratio:
\begin{equation}
 \mathrm{Mg}/\mathrm{Fe}\equiv (x_{\mathrm{Mg}}/x_{\mathrm{Fe}}) \div 
  (x_{\mathrm{Mg},\odot}/x_{\mathrm{Fe},\odot}),  
\end{equation}
where $x_{i,\odot} \equiv M_{i,\odot}/\sum_{i=6}^{30}M_{i,\odot}$ is the
solar metal mass fraction calculated from
\cite{2009ARA&A..47..481A}. Accordingly, note that
$\mathrm{Mg}/\mathrm{Fe}\equiv 10^{[\mathrm{Mg}/\mathrm{Fe}]}$.  Figure
\ref{fig:mgfe_sn_ratio} shows the logarithmic abundance ratio [Mg/Fe] as
a function of $r$.  We calculate the abundance pattern for the specific
values of $r$ corresponding to Mg/Fe=1/10, 1/8, 1/6, 1/4, 1/2, 1, 2, and
4, plus the case of $r=0$ that stands for the pure SN II pattern.
Tables \ref{tab:abuptn_nomoto} and \ref{tab:abuptn_woosley} give the
numerical form of the SN abundance pattern, using the SN II yields of
\cite{1997NuPhA.616...79N} and \cite{1995ApJS..101..181W}, respectively.

\begin{figure}[t]
  \epsscale{1.0}
 \plotone{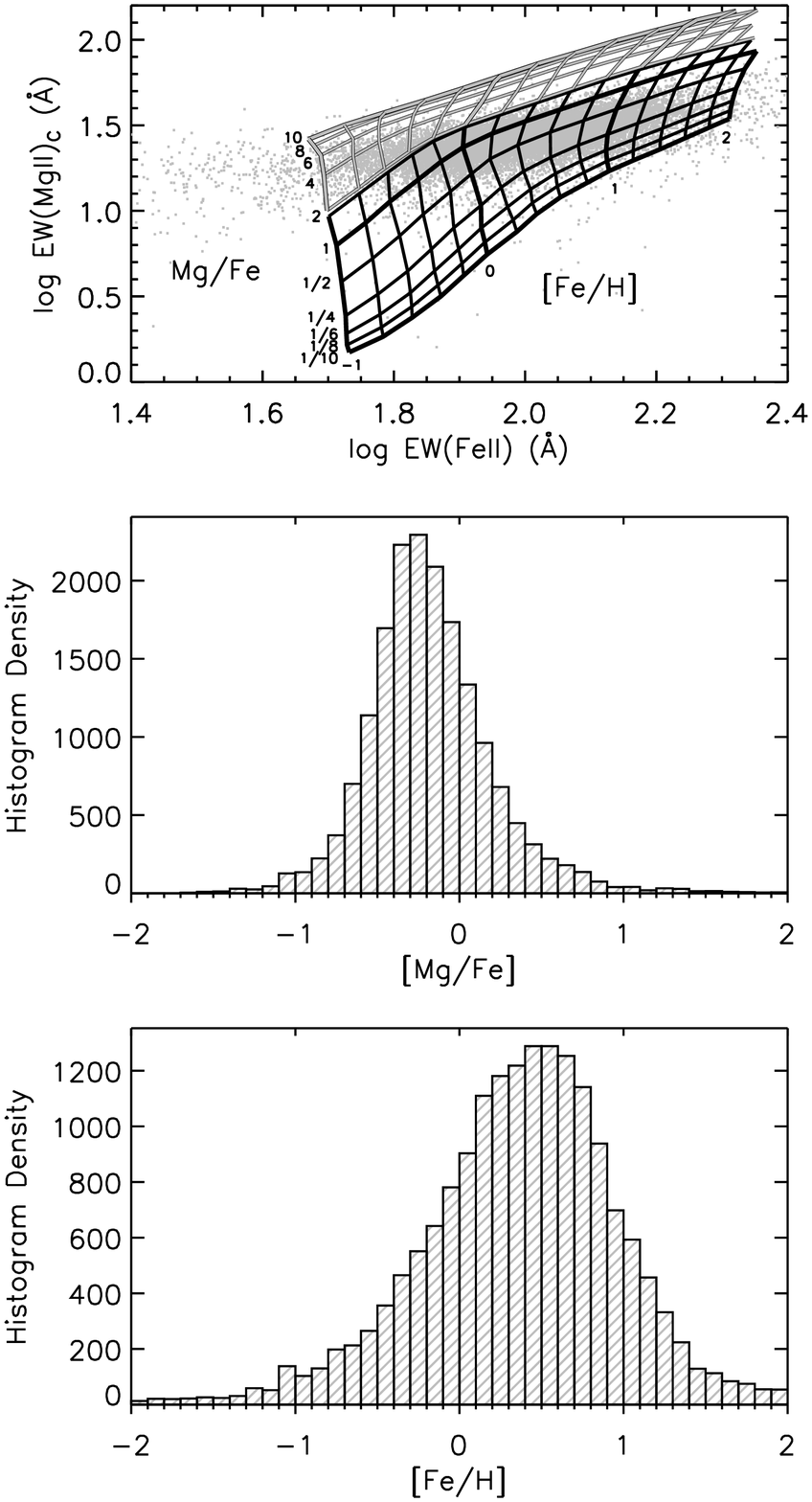}
 \caption{Measurements of Fe and Mg abundances of our sample quasars 
 from \ion{Fe}{2}\ and \ion{Mg}{2}\ emission lines, based on our 
 photoionization calculations using the SN heavy-element abundance pattern  
 (Table \ref{tab:abuptn_nomoto}) from the SN II yields of 
 \cite{1997NuPhA.616...79N}. 
 {\it Top:} The calculated grid of the (Mg/Fe, [Fe/H]) abundance pair  
 overlaid on the EW(\ion{Mg}{2})$_c$ versus EW(\ion{Fe}{2}) diagram. 
 The gray grid is based on additional calculations placing more weight 
 on massive progenitor (see text).  The measured EW(\ion{Mg}{2})$_c$ 
 of our SDSS quasars, which have been corrected for the density effect, 
 are shown by gray dots.  
 {\it Middle:} Histogram of the logarithmic abundance ratio 
 [Mg/Fe] with a bin size of 0.1 dex. {\it Bottom: } Histogram of metallicity 
 [Fe/H] with a bin size of 0.1 dex.}  \label{fig:ew_mgfe_feh_nomoto}
\end{figure}

\begin{figure}[t]
 \epsscale{1.0}
 \plotone{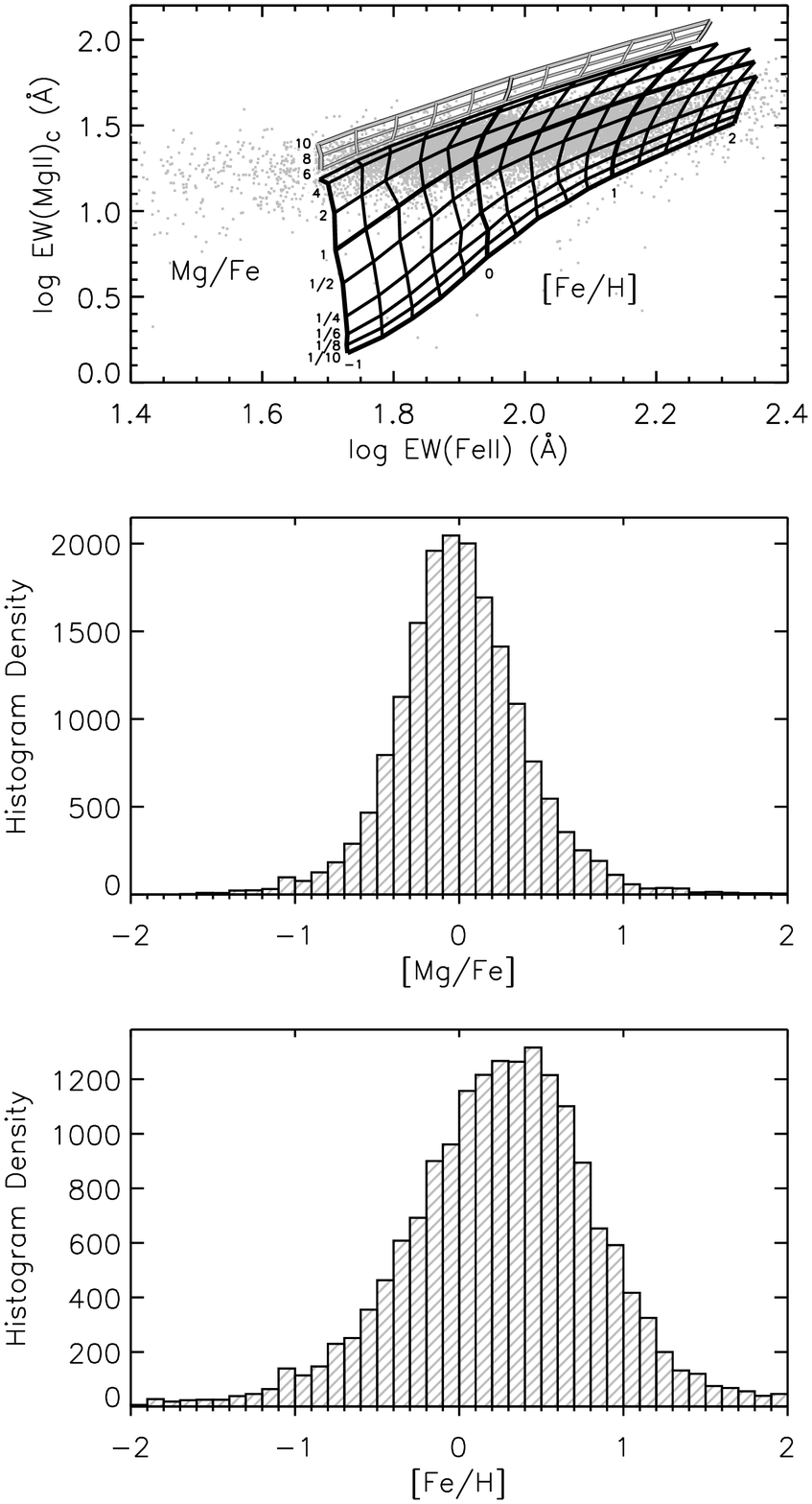}
 \caption{Measurements of Fe and Mg abundances of our sample quasars
 from \ion{Fe}{2}\ and \ion{Mg}{2}\ emission lines.  The same as
 Figure \ref{fig:ew_mgfe_feh_nomoto}, but with the SN heavy-element
 abundance pattern (Table \ref{tab:abuptn_woosley}) from the SN II
 yields of \cite{1995ApJS..101..181W}.}  \label{fig:ew_mgfe_feh_woosley}
\end{figure}

In this study, for each of the $x_i$ patterns having different Mg/Fe
values as above, we run the {\sc Cloudy} code by scaling $x_i$ so that
the resultant grid of Fe/H is from $10^{-1}$ to $10^{+2}$ with a step of
0.2 dex to reflect the metallicity variation, with the shape of the
$x_i$ pattern unchanged. The results of the {\sc Cloudy} calculations
are shown in Figure \ref{fig:ew_mgfe_feh_nomoto} for the case of
\cite{1997NuPhA.616...79N} and Figure \ref{fig:ew_mgfe_feh_woosley} for
the case of \cite{1995ApJS..101..181W}, where the (Mg/Fe, [Fe/H]) grid
is displayed on the corrected EW(\ion{Mg}{2})$_{\mathrm{c}}$ versus
EW(\ion{Fe}{2}) diagram ({\it top panel}). A point to be noted here is
that the {\sc Cloudy} code is run with a covering factor of unity and
the calculated EWs must be adjusted to the observation by scaling the
covering factor of a BLR cloud. We find that the (Mg/Fe, [Fe/H]) grid
with a covering factor of 0.13 nicely covers the measured EWs. This low
value is fully consistent with previous independent reports (e.g.,
\citealt{1997iagn.book.....P}, \citealt{2013peag.book.....N}), and is
therefore used in Figures \ref{fig:ew_mgfe_feh_nomoto} and
\ref{fig:ew_mgfe_feh_woosley}. A small fraction of measured EWs,
however, lie outside the grid corresponding to the maximum Mg/Fe value
expected from the pure SN II yields.  This can be solved by placing more
weight on massive progenitor and extending the grid towards larger Mg/Fe
values. Additional calculations with increasing $m_{\mathrm{min}}$
beyond $10M_\odot$ in equation (\ref{eq:mean_ejecta_mass}) are performed
while fixing $m_{\mathrm{max}}=50M_\odot$. The additional abundance
patterns are tabulated in Tables \ref{tab:abuptn_nomoto} and
\ref{tab:abuptn_woosley}, and the additional grids calculated with these
abundance patterns are shown in gray in Figures
\ref{fig:ew_mgfe_feh_nomoto} and \ref{fig:ew_mgfe_feh_woosley}.

The result that the measured EWs are covered by the (Mg/Fe, [Fe/H]) grid
for reasonable respective ranges of Mg/Fe and [Fe/H] justifies our
conversion of (EW(\ion{Mg}{2})$_{\mathrm{c}}$, EW(\ion{Fe}{2}) to
(Mg/Fe, [Fe/H]).  As can be seen from Figures
\ref{fig:ew_mgfe_feh_nomoto} and \ref{fig:ew_mgfe_feh_woosley}, the
lines of constant Mg/Fe are slightly curved, non-orthogonal to those of
constant [Fe/H], and inclined to the major axis of distribution of the
measured EWs. Therefore, Mg/Fe, as well as [Fe/H], is a non-linear
function of EW(\ion{Mg}{2})$_{\mathrm{c}}$ and EW(\ion{Fe}{2}).  For
reference, Figures \ref{fig:ew_mgfe_feh_nomoto} and
\ref{fig:ew_mgfe_feh_woosley} show the derived distributions of [Mg/Fe]
({\it middle panel}) and [Fe/H] ({\it bottom panel}) for our sample of
SDSS quasars.  Typical uncertainties of the measured [Mg/Fe] and [Fe/H]
are estimated to be 0.27 dex and 0.32 dex, respectively, from the error
propagation of the measurement errors on EW(\ion{Mg}{2}) and
EW(\ion{Fe}{2}).  With no alternative methods reported previously, this
conversion is the first attempt to reliably extract the abundance
information from the emission lines of metals.

\subsection{Evolution of [Mg/Fe] and [Fe/H]}

\begin{figure*}[t]
 \epsscale{1.0}
 \plotone{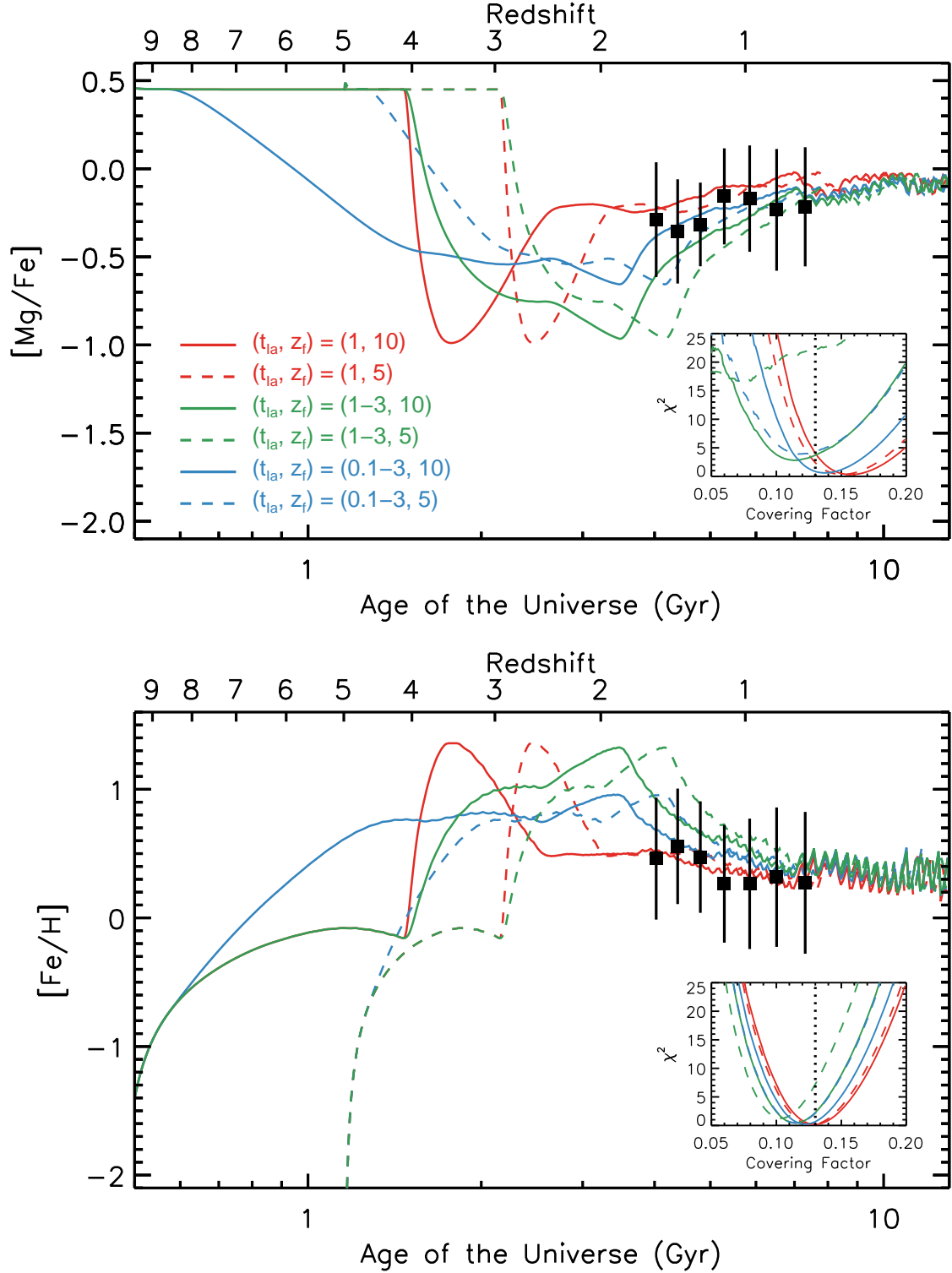}
 \caption{Comparison of various chemical evolution models with derived
 [Mg/Fe] abundance ratios ({\it top panel}) and [Fe/H] abundance ratios
 ({\it bottom panel}) as a function of redshift for our sample of SDSS
 quasars. The mean values and $\sqrt{\sigma^2 -
 \sigma_{\mathrm{mes}}^2}$ of the derived [Mg/Fe] and [Fe/H] abundance
 ratios in each redshift bin ($\Delta z=0.15$) are indicated by filled
 boxes and error bars, respectively. The overplotted curves are the
 chemical evolution models of quasar host galaxies for several
 prescriptions of the SN Ia lifetime and the star formation epoch.  Note
 that the jagged pattern at $z \lesssim 1$ in each model is artificial
 due to the finite time steps in the model calculations.  In the inset,
 a chi-squared value is plotted as a function of the covering factor for
 each of the chemical evolution models, and the adopted covering factor
 of 0.13 in this study is indicated by a vertical dotted line.  Note
 that the SN II yields of \cite{1997NuPhA.616...79N} are used to derive
 [Mg/Fe] and [Fe/H].}  \label{fig:chem_evo}
\end{figure*}

The goal of this study is to constrain the star formation history in the
Universe from the chemical evolution of quasar host galaxies.  In
contrast with previous studies, in which the \ion{Mg}{2}/\ion{Fe}{2}
flux ratio is used as the Mg/Fe abundance ratio, our proposed method
allows measurement of the Mg/Fe abundance ratio, enabling direct
comparison with theoretical chemical evolution models.  Figure
\ref{fig:chem_evo} ({\it top panel}) shows the result of [Mg/Fe] as a
function of redshift for our SDSS quasars plotted in Figure
\ref{fig:ew_mgfe_feh_nomoto}.  The mean values of the derived [Mg/Fe]
abundance ratios in each redshift bin ($\Delta z=0.15$) are indicated by
filled boxes.  The error bars indicate $\sqrt{\sigma^2 -
\sigma_{\mathrm{mes}}^2}$, where $\sigma$ is the standard deviation of
the derived [Mg/Fe] abundance ratios in each redshift bin and
$\sigma_{\mathrm{mes}}$ is the measurement error of [Mg/Fe] estimated in
the previous subsection.  Therefore, they represent systematic errors of
the covering factor and the parameters listed in Table
\ref{tab:baseline}.  Given that a lot of parameters are fixed to the
fiducial values, the derived error bars are so small that support the
validity of our assumptions.  The SN II yields of
\cite{1997NuPhA.616...79N} are used in the figure, but it should be
noted that even if those of \cite{1995ApJS..101..181W} are used, the
result remains almost the same except for a small offset in [Mg/Fe].
Figure \ref{fig:chem_evo} ({\it bottom panel}) shows the result of
[Fe/H] as a function of redshift, in the same way as [Mg/Fe].

In the figure, shown by curves are the chemical evolution models of
quasar host galaxies with an initial burst of star formation and the
time-invariant Salpeter IMF (for details, see
\citealt{1998ApJ...507L.113Y}).  Three types of the SN Ia lifetime are
considered: a single lifetime model with $t_{\mathrm{Ia}}=1$ Gyr, a
box-shaped $t_{\mathrm{Ia}}$ distribution model with $1$--$3$ Gyr
(\citealt{1996ApJ...462..266Y}), and a power-law $t_{\mathrm{Ia}}$
distribution model with $g(t_{\mathrm{Ia}}) \propto
t_{\mathrm{Ia}}^{-0.1}$ for $0.1$--$3$ Gyr
(\citealt{2008PASJ...60.1327T}).  For each of these three types, two
values of the initial star formation redshift are considered: $z_f=5$
and 10.  In general, the [Mg/Fe] ratio is initially maintained at a
level as high as +0.5 dex, reflecting the [Mg/Fe] ratio of the SN II
ejecta averaged with the time-invariant Salpeter IMF, and then the onset
of SNe Ia with the enhanced Fe supply causes a sudden fall and rise of
theoretical curves of [Mg/Fe] and [Fe/H], respectively. After SNe Ia
have ended their lives, a significant number of metal-poor
intermediate-mass stars, which were formed initially from the
metal-deficient gas with high [Mg/Fe] of pure SN II origin, start to
lose their envelope mass in the the surrounding gas.  This can be seen
as upward turn of [Mg/Fe] and downward turn of [Fe/H] at $z\sim 2-3$.
Since the effect of mass loss becomes small with $z$, the theoretical
curves of [Mg/Fe] and [Fe/H] eventually reach their respective constant
levels at $z<1$.

The most important thing to note from Figure \ref{fig:chem_evo} is that
our measurements of [Mg/Fe] and [Fe/H] at redshift below $z <2$ lie
along the theoretical curves. This means that unless the effect of later
mass loss is taken into account in the models, the theoretical curves
will continue to fall in [Mg/Fe] and rise in [Fe/H] with decreasing $z$
and become unable to agree with our measurements at $z < 2$. Therefore,
the result in Figure \ref{fig:chem_evo} strongly suggests that the mass
loss of intermediate-mass stars did occur in quasars.

Another point to note is that the predictions at $z>2$ largely differ
from model to model, although they tend to converge at $z < 2$.  From
the inset diagram of Figure \ref{fig:chem_evo} ({\it top panel}), we may
safely say that the three models adopting $(t_{\mathrm{Ia}},
z_f)$=(1--3, 5), (1--3, 10) and (0.1--3, 5) are rejected because
$\chi^2$ values of those models are clearly larger than the other
models.  However, the convergence of models at low redshift prevents us
from distinguishing which is the best model in the remaining three
models from our sample of SDSS quasars.  For this reason, high-redshift
data are vitally important to overcome the indistinguishability among
the models at $z<2$.  In the forthcoming paper, we will give a more
complete discussion on the chemical evolution of the Universe by also
analyzing the available \ion{Fe}{2}/\ion{Mg}{2} data at $z > 2$.

Comparing our result with chemical evolution studies using other kind of
objects, such as damped Lyman-$\alpha$ (DLA) absorbers, is also of
interest.  As can be seen from Figure 6 in \cite{2012ApJ...755...89R},
the metallicity of DLA is sub-solar and decreases with increasing
redshift, both of which are quite different from our result.  However,
it is not surprising given that BLR clouds and DLA absorbers are totally
different objects; the former likely reside in elliptical galaxies,
while the latter are likely associated with star-forming galaxies.  The
difference between their host galaxies and star formation histories may
account for the different behavior of chemical evolution.  On the other
hand, several studies indicate that sub-DLAs appear to be metal rich.
For example, Figure 11 in \cite{2015ApJ...806...25S} indicates that the
sub-DLA metallicity is super-solar at $z < 1$, which is consistent with
our result.  It is interesting to note that \cite{2013ApJ...775..119C}
observed 21 quasars, which are known to contain sub-DLAs in their
spectra, with the {\it Chandra} X-ray observatory.  They found possible
X-ray emission within $\sim1\arcsec$ of the background quasar in six
cases.  Both of these results suggest that sub-DLAs may be associated
with AGNs.  Further investigation on this aspect would give insights
into galaxy evolution.

It remains, however, a challenge for future research to investigate the
systematic errors of our new abundance diagnostics. In particular, the
assumptions worthy of mention for the fiducial quantities of a typical
BLR cloud include the gas density of $10^{11}\ \mathrm{cm^{-3}}$ and the
covering factor of 0.13. The fiducial gas density determines an anchor
point of the grid shown in Figures 14 and 15 in a vertical direction,
resulting in a systematic error for [Mg/Fe]. On the other hand, the
fiducial covering factor determines an anchor point of the grid in a
diagonal direction, resulting in systematic errors for both [Mg/Fe] and
[Fe/H]. It should be noted, however, that the covering factor changes
the estimated [Mg/Fe] and [Fe/H] in the opposite direction, and is
uniquely determined by comparing the chemical evolution models at low
redshift. Improving the accuracy of these estimates would require
analysis of other emission lines together with \ion{Fe}{2} and
\ion{Mg}{2}, which would make it possible to more fully investigate the
chemical evolution of the Universe with quasar emission lines.

\subsection{Black hole mass and metallicity}

\begin{figure}[t]
 \epsscale{1.0}
 \plotone{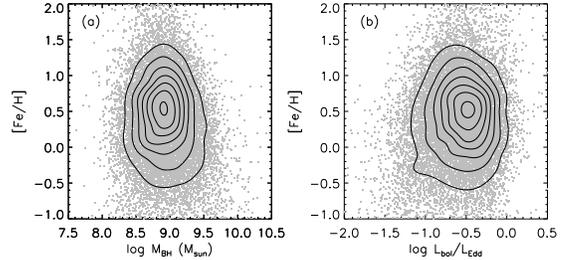}
 \caption{Dependence of metallicity on (a) black hole mass and (b) the
 Eddington ratio. The contours indicate local data point densities with
 a grid size of $\Delta=0.01$ dex.}  \label{fig:bhmass_metal}
\end{figure}

There has been a growing interest in a relationship between black hole
mass and other quasar properties.  We have checked whether the
metallicity, measured by our method, depends on black hole mass or not.
The result is shown in Figure \ref{fig:bhmass_metal}.  Evidently,
apparent correlation is not confirmed.  We also plot the metallicity
against the Eddington ratio, but the result is similar.  Several authors
report the existence of correlations between these quantities and the
\ion{N}{5}/\ion{C}{4} flux ratio, which is frequently used as an
indicator of BLR metallicity (e.g., \citealt{2004ApJ...614..547S},
\citealt{2011A&A...527A.100M}).  Investigating the difference between
their result and ours is beyond the scope of this paper, but it may be
due to the density dependence of \ion{N}{5} and/or \ion{C}{4}, or
different origination between highly ionized lines such as \ion{N}{5}
and \ion{C}{4} and less highly ionized lines such as \ion{Fe}{2}.

\section{Summary and Conclusion}
For the purpose of inventing abundance diagnostics for a BLR cloud, we
selected 17,432 quasars from the SDSS DR7 data and analyzed their
\ion{Mg}{2}\ and \ion{Fe}{2}\ emission lines. In addition to their
emission-line strengths, we measured their black hole masses and
Eddington ratios. As a result, an anticorrelation was found to exist
between the \ion{Mg}{2}/\ion{Fe}{2}\ flux ratio and the Eddington ratio;
this is largely due to the anticorrelation between the \ion{Mg}{2}\ line
strength and the Eddington ratio.

How and where \ion{Mg}{2}\ and \ion{Fe}{2}\ emission lines are created
in a BLR cloud was investigated by photoionization calculations using
{\sc Cloudy}. Taking into account the results of recent reverberation
mapping studies, our calculations show that \ion{Mg}{2}\ and
\ion{Fe}{2}\ are created in different regions of a BLR cloud. This
indicates that the \ion{Mg}{2}/\ion{Fe}{2}\ flux ratio depends not only
on the Mg/Fe abundance ratio but also on other non-abundance
parameters. Therefore, the \ion{Mg}{2}/\ion{Fe}{2}\ flux ratio corrected
for the non-abundance parameter has to be used for the abundance
diagnostics.

In order to identify the main cause that accounts for the observed
trends of the \ion{Mg}{2}\ and \ion{Fe}{2}\ emission-line strengths,
extensive calculations were carried out by varying each of the
non-abundance parameters in {\sc Cloudy}, such as BLR gas density,
column density, SED of the incident continuum, and microturbulence. It
was found that the observed trends can be reproduced solely by changing
the BLR gas density. We have thus concluded that the trend of
\ion{Mg}{2}/\ion{Fe}{2}\ with the Eddington ratio for our sample of SDSS
quasars requires the existence of anticorrelation between the BLR gas
density and the Eddington ratio.

Accordingly, \ion{Mg}{2}/\ion{Fe}{2}\ corrected for the density effect
should reflect the abundance ratio more directly. To confirm this
expectation, the corrected EW(\ion{Mg}{2})$_{\mathrm{c}}$ versus
EW(\ion{Fe}{2}) diagram is compared with calculations from {\sc Cloudy},
where the Mg/Fe abundance ratio and the metallicity [Fe/H] are varied
with other non-abundance parameters fixed to their fiducial values for
the baseline model of a BLR cloud.  It is found that the data in the
diagram are distributed within the calculated grid of Mg/Fe and [Fe/H]
for their reasonable respective ranges.  This indicates that our
abundance diagnostics works well for a BLR cloud, and more reliable
estimates of the gas density and covering factor for a BLR cloud further
enhance the accuracy of the resultant estimates of abundance parameter
values.

For our sample of SDSS quasars at $z<2$, we have derived
$[\mathrm{Mg/Fe}] \sim -0.2$ and $[\mathrm{Fe/H}] \sim +0.5$ on the
average, which more or less agrees with chemical evolution models in
which the effect of mass loss from initially formed metal-poor
intermediate-mass stars is explicitly taken into account.  These models,
although tending to converge at $z<2$, predict the diversity in [Mg/Fe]
at $z>2$. Therefore, high-redshift data are essential for distinguishing
the best model from many plausible ones at $z<2$.  We will discuss this
in the forthcoming paper.

Finally, we have compared the derived [Fe/H] with black hole mass and
the Eddington ratio.  Contrary to the earlier studies using the
\ion{N}{5}/\ion{C}{4} flux ratio as an indicator of BLR metallicity, no
apparent correlation between them is confirmed.  Although investing the
origin of the difference is beyond the scope of this paper, it may be
due to the density dependence of \ion{N}{5} and/or \ion{C}{4}, or
different origination between highly ionized lines such as \ion{N}{5}
and \ion{C}{4} and less highly ionized lines such as \ion{Fe}{2}.


\acknowledgments

We acknowledge the anonymous referee for constructive comments that
helped to improve the quality of our manuscript.  We thank Takuji
Tsujimoto and Toshihiro Kawaguchi for helpful discussions. This work was
supported by Grant-in-Aid for JSPS Fellows Grant Number 12J10755, and
JSPS KAKENHI Grant Number 22253002.  Funding for the SDSS and SDSS-II
has been provided by the Alfred P. Sloan Foundation, the Participating
Institutions, the National Science Foundation, the U.S. Department of
Energy, the National Aeronautics and Space Administration, the Japanese
Monbukagakusho, the Max Planck Society, and the Higher Education Funding
Council for England. The SDSS Web Site is http://www.sdss.org/. The SDSS
is managed by the Astrophysical Research Consortium for the
Participating Institutions. The Participating Institutions are the
American Museum of Natural History, Astrophysical Institute Potsdam,
University of Basel, University of Cambridge, Case Western Reserve
University, University of Chicago, Drexel University, Fermilab, the
Institute for Advanced Study, the Japan Participation Group, Johns
Hopkins University, the Joint Institute for Nuclear Astrophysics, the
Kavli Institute for Particle Astrophysics and Cosmology, the Korean
Scientist Group, the Chinese Academy of Sciences (LAMOST), Los Alamos
National Laboratory, the Max-Planck-Institute for Astronomy (MPIA), the
Max-Planck-Institute for Astrophysics (MPA), New Mexico State
University, Ohio State University, University of Pittsburgh, University
of Portsmouth, Princeton University, the United States Naval
Observatory, and the University of Washington.


\end{document}